\documentclass[letterpaper]{article}

\title{The Power of Migrations in Dynamic Bin Packing} 

\author{Anonymous authors}{ }{}{}{}
\date{}

\usepackage{xcolor}
\usepackage{graphicx} 
\usepackage{amsmath, amsthm, amssymb}
\usepackage{url}
\usepackage{bm}
\usepackage{fullpage}
\usepackage[procnumbered, linesnumbered,algoruled]{algorithm2e}
\usepackage{xurl}
\usepackage{thmtools,thm-restate}
\usepackage[hidelinks]{hyperref}

\usepackage{cleveref}
\usepackage{comment}

\newcommand{\red}[1]{{\color{red} #1}}
\newcommand{\blue}[1]{{\color{blue} #1}}
\newcommand{\e}{\varepsilon}
\newcommand{\cI}{\mathcal{I}}
\newcommand{\OPT}{\mathrm{OPT}}
\newcommand{\ALG}{\mathrm{ALG}}
\newcommand{\E}[0]{\ensuremath \mathbb{E}}
\newcommand{\Es}[0]{\ensuremath \mathbb{E}\,}
\newcommand{\cA}{\mathcal{A}}

\newcommand{\ones}{\bm{1}}
\newcommand{\cX}{\mathcal{X}}
\newcommand{\Y}{\mathcal{Y}}
\newcommand{\mig}{\textup{mig}}
\newcommand{\cost}{\textup{cost}}
\newcommand{\full}{\textup{full}}
\renewcommand{\d}{\textup{d}}
\newcommand{\R}{\mathcal{R}}
\newcommand{\act}{\rho}
\newcommand{\vol}{\textup{Vol}}
\newcommand{\spa}{\textup{Span}}

\newcommand{\dyn}{\textsc{DynBinPack}\xspace}
\newcommand{\dynd}{\textsc{DynBinPack-Delay}\xspace}

\newcommand{\remove}[1]{}

\newtheorem{lemma}{Lemma}
\newtheorem{thm}{Theorem}
\newtheorem{cor}{Corollary}
\newtheorem{claim}{Claim}

\definecolor{lavender}{rgb}{0.9, 0.9, 0.98}

\newcounter{mynotes}
\newcommand{\knote}[1]{\addtocounter{mynotes}{1}{{\bf !}}%
\marginpar{\scriptsize  {\arabic{mynotes}.\ {\sf \textcolor{red}{#1}}}}}

\newcommand{\rnote}[1]{\addtocounter{mynotes}{1}{{\bf !}}%
\marginpar{\scriptsize  {\arabic{mynotes}.\ {\sf \textcolor{blue}{#1}}}}}

\newcommand{\mnote}[1]{\addtocounter{mynotes}{1}{{\bf !}}%
\marginpar{\scriptsize  {\arabic{mynotes}.\ {\sf \textcolor{magenta}{#1}}}}}

\renewcommand{\red}[1]{#1}
\renewcommand{\blue}[1]{#1}
\renewcommand{\knote}[1]{}
\renewcommand{\rnote}[1]{}
\renewcommand{\mnote}[1]{}

\usepackage[normalem]{ulem}

\usepackage{tikz}
\usetikzlibrary{arrows,positioning}
\usetikzlibrary{patterns}
\usetikzlibrary{decorations.pathreplacing}
\usetikzlibrary{calc}
\usepackage[outline]{contour}
\contourlength{1.2pt}

\newcommand{\firstfit}{\textsc{FirstFit}\xspace}

\author{Konstantina Mellou \\ Microsoft Research Redmond \\ kmellou@microsoft.com
	\and Marco Molinaro \\ Microsoft Research Redmond and PUC-Rio \\ mmolinaro@microsoft.com
	\and Rudy Zhou\thanks{Supported in part by ONR Award N000142212702} \\ Carnegie Mellon University \\ rbz@andrew.cmu.edu}

\begin{document}

\maketitle

\begin{abstract}
In the \emph{Dynamic Bin Packing} problem, $n$ items arrive and depart the system in an online manner, and the goal is to maintain a good packing throughout. We consider the objective of minimizing the total active time, i.e., the sum of the number of open bins over all times. \blue{An important tool for maintaining an efficient packing in many applications is the use of \emph{migrations}; e.g., transferring computing jobs across different machines.}
However, there are large gaps in our understanding of the approximability of dynamic bin packing with migrations. Prior work has covered the power of no migrations and $> n$ migrations, but we ask the question: What is the power of limited ($\leq n$) migrations?

Our first result is a dichotomy between no migrations and linear migrations: Using a sublinear number of migrations is asymptotically equivalent to doing \emph{zero} migrations, where the competitve ratio grows with $\mu$, the ratio of the largest to smallest item duration. On the other hand, we prove that for every $\alpha \in (0,1]$, there is an algorithm that does $\approx \alpha n$ migrations and achieves competitive ratio $\approx 1/\alpha$ (in particular, independent of $\mu$); we also show that this tradeoff is essentially best possible. This fills in the gap between zero migrations and $> n$ migrations in Dynamic Bin Packing. 

Finally, in light of the above impossibility results, we introduce a new model that more directly captures the impact of migrations. Instead of limiting the number of migrations, each migration adds a delay of $C$ time units to the item's duration; \blue{this commonly appears in settings where a blackout or set-up time is required before the item can restart its execution in the new bin.}
In this new model, we prove a $O(\min (\sqrt{C}, \mu))$-approximation, and an almost matching lower bound.
\end{abstract}

\thispagestyle{empty}
\clearpage

\setcounter{page}{1}

\newpage 

\section{Introduction}


We study the problem of \emph{online (fully) dynamic Bin Packing} (\dyn) with migrations. In this version of the classical Bin Packing problem, items both enter and leave the system. Without migrations, this problem has been extensively studied since the 80's~\cite{coffmanGJ,AzarVainstein,nonClairvoyantDynBP,ivkovic1998fully,feldkord2018fully,tang2016first,balogh2008lower,luo,jansen,berndt}. In this setting,
we maintain a collection of unit-sized bins. Items arrive online, at their \emph{arrival time} $a_i \geq 0$, with \emph{size}  $s_i \in [0,1]$ and \emph{duration} $d_i > 0$ for item $i = 1,\ldots,n$. Upon arrival, an item must be either assigned to an open bin with enough space to accommodate the item size, or we must open a new bin and place the item there. After duration-many time units pass, the item departs from the system. Since \blue{in many applications (e.g., cloud job scheduling),} the duration of an item is determined by the user, we consider the \emph{non-clairvoyant} model, where upon item arrival only its size is revealed, but its duration is unknown.
The goal is to minimize the \emph{total active time} of the bins, namely $\sum_{i} \int_{t = 0}^\infty \ones(\text{bin $i$ open at time $t$})\,\d t$. 
As usual, we compare against the optimal offline solution $\OPT$ that knows all items upfront and can arbitrarily re-pack items at any time.




A significant additional dimension to this problem is the use of \emph{migrations}, 
i.e., the ability of moving an already placed item to a different bin, to improve the packing.  
Migrations play a key role in various resource management problems; see~\cite{hadary2020protean} for how cloud providers rely on migrations to maintain efficient placements under load fluctuations, as an example. However, performing a migration often means introducing additional overhead to the system. That is the result of migrations requiring a ``blackout'' or ``set-up'' time for the migration to be completed and the item to resume its execution on the new bin~\cite{ruprecht2018vm}; this can eventually cause delays in the completion of the item (e.g., if each item corresponds to a simulation/batch job, see \Cref{fig:delay-migration} for an illustration).   

\vspace{0.1cm}








\begin{figure}[h]
    \centering
    \begin{minipage}[t]{.34\textwidth}
    \centering
    \scalebox{0.45}{
\begin{tikzpicture}

\node[black,right] at (3.7, 8.3) {\huge No migrations:};
\draw (4,6.2)[fill=lavender] rectangle (11,6.7); 
\node[black] at (4.2, 5.7) {\LARGE $a_0$};
\node[black] at (11.5, 5.7) {\LARGE $a_0+d_0$};
\draw [decorate,
    decoration = {brace,mirror}] (11,6.8) --  (4,6.8);
\node[black] at (7.5,7.3) {\LARGE Bin 1};


\node[black,right] at (3.7, 4.1) {\huge 2 migrations: };

\draw (4,1.8)[fill=lavender] rectangle (13,2.3); 
\draw (6,1.8)[fill=black] rectangle (7,2.3); 
\node[black] at (4.2, 1.3) {\LARGE $a_0$};
\node[black] at (13, 1.3) {\LARGE $a_0+d_0+2C$};
\node[black] at (6.5,1.3) {\LARGE Delay $C$};


\draw [decorate,
    decoration = {brace,mirror}] (5.9,2.4) --  (4,2.4);
\node[black] at (5,2.9) {\LARGE Bin 1};
\draw [decorate,
    decoration = {brace,mirror}] (9.4,2.4) --  (6,2.4);
\node[black] at (7.7,2.9) {\LARGE Bin 2};
\draw [decorate,
    decoration = {brace,mirror}] (13,2.4) --  (9.5,2.4);
\node[black] at (11.2,2.9) {\LARGE Bin 3};

\tikzset{invisible/.style={minimum width=0mm,inner sep=0mm,outer sep=0mm}}
\node[invisible, white] at (6, 1) {$\frac{1}{4}$};

\draw (9.5,1.8)[fill=black] rectangle (10.5,2.3); 
\node[black] at (10,1.3) {\LARGE Delay $C$};
\end{tikzpicture}
}
\caption{Illustration of an item with arrival time $a_0$ and duration $d_0$ with 0 and 2 migrations. Each migration leads to a delay of length $C$ in its total duration. \medskip 
}
\label{fig:delay-migration}
    \end{minipage}
\hfill
    \begin{minipage}[t]{.61\textwidth}
    \centering
 \scalebox{0.5}{
\begin{tikzpicture}
\draw (0,0) rectangle (7,1.25); 
\draw (0,0)[fill=lavender] rectangle (7,0.25); 
\draw (0,0.25)[fill=lavender] rectangle (0.25,0.5); 
\draw (0,0.5)[fill=lavender] rectangle (0.25,0.75); 
\draw (0,0.75)[fill=lavender] rectangle (0.25,1,0); 
\draw (0,1.0)[fill=lavender] rectangle (0.25,1.25); 

\draw (0,1.75) rectangle (7,3); 
\draw (0,1.75)[fill=lavender] rectangle (7,2); 
\draw (0,2.)[fill=lavender] rectangle (0.25,2.25); 
\draw (0,2.25)[fill=lavender] rectangle (0.25,2.5); 
\draw (0,2.5)[fill=lavender] rectangle (0.25,2.75); 
\draw (0,2.75)[fill=lavender] rectangle (0.25,3); 

\node[black] at (3.5, 3.85) {\large $\vdots$};

\draw (0,4.5) rectangle (7,5.75); 
\draw (0,4.5)[fill=lavender] rectangle (7,4.75); 
\draw (0,4.75)[fill=lavender] rectangle (0.25,5); 
\draw (0,5.)[fill=lavender] rectangle (0.25,5.25); 
\draw (0,5.25)[fill=lavender] rectangle (0.25,5.5); 
\draw (0,5.5)[fill=lavender] rectangle (0.25,5.75); 
\node[black] at (3.5, -0.8) {\LARGE Bad allocation};
\draw (9,0) rectangle (16,1.25); 
\draw (9,0)[fill=lavender] rectangle (16,0.25); 
\draw (9,0.25)[fill=lavender] rectangle (16,0.5); 
\draw (9,0.5)[fill=lavender] rectangle (16,0.75); 
\draw (9,0.75)[fill=lavender] rectangle (16,1,0); 
\draw (9,1.0)[fill=lavender] rectangle (16,1.25); 

\draw (9,1.75) rectangle (9.25,3); 
\draw (9,1.75)[fill=lavender] rectangle (9.25,2); 
\draw (9,2.)[fill=lavender] rectangle (9.25,2.25); 
\draw (9,2.25)[fill=lavender] rectangle (9.25,2.5); 
\draw (9,2.5)[fill=lavender] rectangle (9.25,2.75); 
\draw (9,2.75)[fill=lavender] rectangle (9.25,3); 

\node[black] at (9.12, 3.85) {\large $\vdots$};

\draw (9,4.5) rectangle (9.25,5.75); 
\draw (9,4.5)[fill=lavender] rectangle (9.25,4.75); 
\draw (9,4.75)[fill=lavender] rectangle (9.25,5); 
\draw (9,5.)[fill=lavender] rectangle (9.25,5.25); 
\draw (9,5.25)[fill=lavender] rectangle (9.25,5.5); 
\draw (9,5.5)[fill=lavender] rectangle (9.25,5.75); 
\node[black] at (12.5, -0.8) {\LARGE Optimal allocation};
\end{tikzpicture}
}
\caption{$k^2$ items of size $\frac{1}{k}$ arrive at time 0. The algorithm puts them in $k$ bins without knowing their duration. The adversary picks one item on each bin to be long-lived (duration $\mu$) and all others  short-lived (duration 1). The cost of the algorithm is $k \cdot \mu$. An optimal allocation puts all long-lived items on one bin and has cost $\mu + (k-1)$. The competitive ratio $\frac{k \cdot \mu}{\mu + (k-1)}$ goes to $\mu$ as $k\rightarrow \infty$.}
\label{fig:bad_example}
    \end{minipage}
\end{figure}

Dynamic Bin Packing with limited migrations has been extensively studied~\cite{IVKOVIC1996229,gambosi,BaloghBGM09,balogh2008lower,BaloghBGR14,feldkord2018fully,epsteinLevinSize,jansen,berndt,feldkord2018fully}. Despite the practical importance and all the attention received, there are still big gaps in the theoretical foundations of the power of limited amounts of migrations. To simplify the discussion, we focus for now on the most natural model that measures the amount of migrations as the \emph{number} of items migrated (so-called \emph{unit-cost} model~\cite{IVKOVIC1996229,gambosi,BaloghBGM09,balogh2008lower,BaloghBGR14,feldkord2018fully}). On one hand, in \dyn with \emph{no migrations} allowed, 
it is known that $\Theta(\mu)$ is the best competitive ratio achievable, where $\mu$ is the ratio of the longest to shortest item duration~\cite{nonClairvoyantDynBP}; see \Cref{fig:bad_example} for an informal presentation of the lower bound. In practice the largest and smallest durations can be orders of magnitude apart, so this guarantee is unsatisfactory. 
On the other hand, it is known that when a \emph{linear} number of migrations is used, the competitive ratio can be greatly improved. More precisely, after a sequence of works,~\cite{feldkord2018fully} designed an algorithm with competitive ratio essentially $1.3871 + \e$ while using $\frac{n}{\e^2}$ migrations, for $\e \in (0,1)$. They actually consider a harder \emph{per-time} objective, ensuring that at all times $t$, the number of bins opened by the algorithm is at most $(1.3871 + \e) \cdot \OPT_t + O(\frac{1}{\e})$, where $\OPT_t$ is the minimum number of bins required for packing the items in the system at time $t$. However, migrating each item on average $\frac{1}{\e^2}$-times is prohibitive in practice due to their impact on the users.
To the best of our knowledge, nothing is known for the most practically relevant regime that lies between these two extremes. Thus, the first question we raise in this work is the following:


\begin{quote}
	\emph{What competitive ratio can be achieved for online dynamic bin packing with $\leq n$ migrations?}
\end{quote}

Given the tight control on the quantity of migrations in practice, it would be of particular interest to obtain a smooth tradeoff between the amount of migrations allowed (e.g., $1\%$ of the items) and the competitive ratio that can be achieved.

Next, we observe that the amount of migrations is a proxy for the operational costs associated with migration, such as the extra overhead in the system that migrations generate, as discussed above. Thus, we propose to shift the focus from the amount of migrations, which has been considered so far, to designing algorithms that consider the actual cost of migrations more directly. Stated informally:


\begin{quote}
 \emph{What competitive ratio can be achieved for online dynamic bin packing where each migration introduces a delay on the item?}
\end{quote}

\subsection{Our Results}


In this paper, we provide quite comprehensive answers to both questions above. For all our results, we compare our algorithms to the \emph{same} optimal policy, $\OPT$, that is allowed to \emph{arbitrarily re-pack} items at any time. We note that this is the standard optimal policy considered in prior works on dynamic bin packing (see, e.g.\cite{coffman1983dynamic, li2015dynamic, feldkord2018fully}.)

\medskip
\noindent\textbf{Dynamic Bin Packing with (sub-) linear amounts of migrations.} Our first result is that the current dichotomy between doing no migrations or a linear number of migrations is actually unavoidable:
a sublinear number of migrations is asymptotically equivalent to doing \emph{zero} migrations. Prior lower bounds for \dyn only cover the regime where the number of migrations is $\Omega(poly(n))$. In particular, \cite{balogh2008lower, feldkord2018fully} show that achieving competitive ratio better than $\approx 1.38$ requires at least $poly(n)$ migrations. However, this does not rule out -- for example, a constant-competitive algorithm using $o(n)$ migrations. Our first result rules out this possibility.


	
\begin{thm}[\Cref{thm:sublinear} (Informal)] \label{thm:subInformal}
	Any randomized algorithm for \dyn that does $o(n)$ migrations in expectation has competitive ratio no better (within a constant factor) than an algorithm that does not do any migrations with respect to the total active time objective. 
\end{thm}

In particular, any algorithm that does $o(n)$ migrations has competitive ratio $\Omega(\mu)$. In fact, our proof strategy goes beyond this specific result and can be used to show that this is actually a broader phenomenon that holds for different variants of the problem, e.g., the clairvoyant version where the online algorithm knows the duration of the item when it arrives, for different migration costs, etc. While somewhat intuitive, the proof turns out to be more technical than expected (as discussed in the technical overview \S \ref{sec:overview}).

Given the limitations revealed by this result, we focus on the regime of using a linear number of migrations and analyze the tradeoff between the fraction of items migrated and the competitive ratio. We essentially show that
for any $\alpha \in (0,1)$, if an $\approx \alpha$-fraction of items can be migrated, then one can achieve competitive ratio $\approx \frac{1}{\alpha}$, and this is best possible. More precisely, we design an algorithm for the stronger objective of maintaining a per-time approximation of the optimal packing as in \cite{feldkord2018fully}; to state the result, recall that $\OPT_t$ denotes the minimum number of bins required for packing the items that are in the system at time $t$. Note that this is exactly the number of bins opened by $\OPT$ at time $t$, since $\OPT$ can arbitrarily repack at any moment.

\begin{restatable}{thm}{unitMigrationCosts} \label{thm:unit-migration-costs}
    For every $\alpha \in (0,\frac{1}{2})$, there is an online algorithm for \dyn that makes at most $\frac{4\alpha}{1-2\alpha} \cdot n$ migrations and for every time $t$ has at most $\frac{1}{\alpha} \OPT_t + O(\log \act)$ open bins, where $\act$ is the maximum number of simultaneous items in the system at any time.
\end{restatable}


Our accompanying lower bound shows that this tradeoff is almost best possible, even for the weaker total active time objective. 


\begin{restatable}{thm}{migrationtradeoff}\label{lem:migrationtradeoff}
	For every $\alpha \in (0,1)$, 
 any randomized algorithm for \dyn that makes at most $\alpha n$ migrations in expectation has competitive ratio at least $\frac{1}{4} \lfloor \frac{1}{4\alpha} \rfloor$ with respect to the total active time objective. 
\end{restatable}

Taken together, our results almost completely characterize the power of migrations for \dyn: Consider any algorithm that does $\alpha n$ migrations.

\begin{itemize}
	\item \textbf{(New)} If $\alpha \rightarrow 0$ as $n \rightarrow \infty$, then the algorithm has no advantage over zero migration algorithms.
	\item \textbf{(New)} If $\alpha \le 1$ is constant, then the correct trade-off
 between the number of migrations and competitive ratio is \emph{multiplicative} -- doing $\ell$-times as many migrations improves the competitive ratio from $\Theta(\frac{1}{\alpha})$ to $\Theta(\frac{1}{\ell \alpha})$.
	\item \textbf{(From \cite{feldkord2018fully})} If $\alpha > 1$, then the correct trade-off is \emph{additive} -- doing $\ell$-times as many migrations decreases the competitive ratio from $1.3871 + O(\frac{1}{\sqrt{\alpha}})$ to $1.3871 + O(\frac{1}{\sqrt{\ell}} \frac{1}{\sqrt{\alpha}})$. Further, improving the competitive ratio requires increasing the number of migrations or additive term in the competitive ratio by a $poly(n)$-factor.
\end{itemize}


In addition to the \emph{unit-cost} migration model considered so far (where the amount of migration is measured by the \emph{number} of items migrated), another popular model in the literature is the \emph{size-cost} migration model, where one counts the total \emph{size} of items migrated instead~\cite{epsteinLevinSize,jansen,berndt,feldkord2018fully}. Again, all known results in this model performed a total size of migrations that was more than the total size of the items (this is analogous to doing $> n$ migrations in the unit-cost model). We prove analogues of Theorems \ref{thm:subInformal}, \ref{thm:unit-migration-costs}, and \ref{lem:migrationtradeoff}, with the same (or tighter) guarantees for this size-cost model as well. Thus, we show that also in this model  $o(\textrm{\it total item sizes})$ migrations (informally) do not offer any power, and prove a smooth competitive ratio tradeoff using $\alpha \cdot \textrm{\it total item sizes}$ (for $\alpha \in [0,1]$) amount of migrations; see \Cref{app:size-migration} for details. 

\begin{thm}[Informal] \label{thm:sizeCost}
	Consider \dyn in the size-cost model. Then, 
 \begin{itemize}
     \item Any randomized algorithm that does $o(\textrm{total item sizes})$ migrations in expectation has competitive ratio no better (within a constant factor) than an algorithm that does not do any migrations. 
     \item For every $\alpha \in (0,\frac{1}{2})$, there is an online algorithm that makes at most $\frac{4\alpha}{1-2\alpha} \cdot \textrm{(total item sizes)}$ migrations and for every time $t$ has at most $\frac{1}{\alpha} \OPT_t + 1$ open bins.
     \item No randomized algorithm that makes  $\alpha \cdot \textrm{(total item sizes)}$ has competitive ratio better than $\frac{1}{4} \lfloor \frac{1}{4\alpha} \rfloor$ with respect to the total active time objective.
 \end{itemize}

\end{thm}




\medskip
\noindent\textbf{Dynamic Bin Packing with delays.} 
Finally, we move on to our second question. Here we introduce the problem of \emph{fully dynamic bin packing with delays} (\dynd): The problem is precisely as \dyn, but every time an item $i$ is migrated, its duration is increased by $+ C$, where the delay cost $C$ is an input parameter known to the algorithm. 

We note that even the general model of \cite{feldkord2018fully} where the algorithm pays a different cost for each migration (instead of paying 1 per migration) and one still wants to have bounded total cost, does not capture the effect of delays: in the latter, making a migration adds additional load to the system, which can impact future assignments. 
In addition, algorithms proposed for the models without delays cannot be easily translated to the delay setting. The main challenge stems from the fact that these low-recourse algorithms make migrations without considering the duration of the items, so they can incur large delays by migrating short items to maintain the per-time objective.
We hope that our new model, \dynd, can be the foundation for further research on dynamic bin packing with migrations that capture additional operational considerations.

Our main result here is the first competitive algorithm for \dynd and an almost-matching lower bound. We again compare with the same optimal policy as before: the optimal policy that is allowed to re-pack arbitrarily, \emph{accruing no delay cost}.

\begin{restatable}{thm}{migrationDelay}\label{thm:alg-delay-cost}
 	For any $C \geq 1$ and assuming the minimum item duration is $1$, there is an online algorithm for \dynd with delay cost $C$ that is $O(\min(\sqrt{C}, \mu))$-competitive, where $\mu$ is the ratio of the longest to shortest item durations.  Moreover, every randomized algorithm has competitive ratio at least $\Omega(\min(\sqrt{C}, \mu))$. 
\end{restatable}


    To put this result in perspective, notice that one can obtain a $O(\mu)$-competitive solution for \dynd by applying the best online algorithm for \dyn that makes no migrations. In \blue{cloud computing applications}, the minimum item duration is in the order of milliseconds, and several last more than years~\cite{hadary2020protean}, so $\mu$ is of the order $\approx 10^{10}$. In contrast, migration delays are rarely more than a few seconds, and so $\sqrt{C}$ is rarely over $10^2$. Thus, the guarantee of \Cref{thm:alg-delay-cost} in practice can be orders of magnitude that of the best algorithm without migrations (the constants in the $O(\cdot)$ are comparable in both results).


		

\subsection{Technical Overview} \label{sec:overview}

For simplicity, the technical overview focuses on the unit-cost model; similar results for the size-cost model can be found in Appendix \ref{app:size-migration}.


\medskip 
\noindent \textbf{Sublinear migrations do not help (\Cref{sec:sublinear}).} To prove the limited power of sublinear migrations, we use, at a high-level, the natural idea of ``repeating a hard instance multiple times to force the need of a linear number of migrations''. However, formalizing this turns out to be a bit more technical than expected.

More precisely, we show 
that from every $\beta$-competitive randomized algorithm $\cA_R$ with sublinear migrations we can extract an algorithm $\cA'_R$ with no migrations that is $\Theta(\beta)$ competitive; the last property can be equivalently stated as  
	\begin{align}
		&\textrm{\textbf{there is} an algorithm $\cA'_R$ with no migrations} \notag\\[-5pt]
		&~~~~\qquad\qquad\textrm{such that \textbf{for every} instance distribution $\nu$, } \E_{I \sim \nu,R} \frac{\cost(\cA'_R,I)}{\OPT(I)} \le \Theta(\beta) \label{eq:mainLB}
	\end{align}
	Unfortunately we cannot show this directly. What we can prove is a much weaker statement (\Cref{lemma:tensor}) where: 
	\begin{enumerate}
		\item The quantifiers are exchanged, namely we show that ``\textbf{for every} distribution $\nu$, \textbf{there is} an algorithm $\cA'_R$\ldots'', that is, the algorithm can be tailored to a known instance distribution.
		
		\item The guarantee of the algorithm $\cA'_R$ bounds the \emph{ratio of expectations} $\frac{\E_{I \sim\nu, R} \cost(\cA'_R, I)}{\E_{I \sim \nu} \OPT(I)}$; however, what we need is a bound on the \emph{expected ratio} $\E\, \frac{\cost(\cA'_R,I)}{\OPT(I)}$. Notice that for general random variables $X$ and $Y$, the  desired \emph{expected ratio} $\E\, \frac{X}{Y}$ can be arbitrarily larger than $\frac{\E X}{\E Y}$. 
  For example, if $X$ is a Bernoulli random variable taking values $0$ and $1$ with equal probability and $Y = 1- X$, we see that $\frac{\E X}{\E Y} = 1$ but $\E \frac{X}{Y} = \infty$.
	\end{enumerate} 
    The proof of \Cref{lemma:tensor} is indeed based on a ``tensoring'' argument, where we run the algorithm $\cA_R$ on a concatenation of instances $I_1,I_2\,\ldots$ sampled from $\nu$ and show that there exists an index $j$ where in the $j$-th part of the instance the algorithm has small cost and makes a tiny amount of migrations (which can be ignored while essentially not changing the cost). By feeding simulated instances $\tilde{I}_1,\ldots,\tilde{I}_{j-1} \sim \nu$ to $\cA_R$ and then running it on an actual instance $I$, we obtain an algorithm for solving $I$ with small cost and no migrations, i.e., the desired algorithm $\cA'_R$.
    
    \blue{To remove the two limitations of this result described in the items above, we give a minimax theorem (\Cref{thm:extendedMinimax}) that allows us to relate the ``expected ratio'' to ``ratio of expectations.'' The quantifier exchange needed in Item 1 is the basis of minimax theorems, so it is a natural tool for that. However, the standard Yao-type minimax theorems do not explicitly involve ratios, i.e., they are of the form (informally) $\min_{\textrm{rand algo $\cA_R$}}$
    $\max_{\textrm{instance $I$}} \E_R f(\cA_R, I)$
    $= \max_{\textrm{rand instance $I$}} \min_{\textrm{algo $\cA$}} \E_I f(\cA, I)$. In particular, they cannot convert from ``expected ratio'' to ``ratio of expectations'' as needed in Item 2. Our \Cref{thm:extendedMinimax} does this and, perhaps surprisingly, \knote{Is a verb missing before "that" here?} we show that in a minimax sense the ``expected competitive ratio'' and ``competitive ratio of expectations'' are the same, even though this is definitely not the case on a per algorithm/instance distribution basis, as the example above shows.
    
    We note that a recent paper~\cite{DBLP:journals/corr/abs-2311-17038} proves a similar general minimax result to ours (\Cref{thm:extendedMinimax}), albeit with a different proof and slightly different setting. Our result allows both ``players'' to randomize, while theirs only allows the ``max player'' to randomize. Further, our proof is a quick consequence of a recent minimax result of Ben-David and Blais~\cite{minimaxBlais}, while they give a direct, elementary proof. Another recent paper \cite{stefanoRatio} also explores such an equivalence between ratio-of-expectation and expectation-of-ratio by designing an explicit algorithm (in the context of prophet inequalities.)}

    Lastly, while it is possible to obtain a more streamlined proof of the limitation of sublinear migrations in \dyn by focusing on specific hard instances for this problem, we believe our proof via the ``transfer'' Theorem \ref{thm:subInformal} is valuable for two reasons: 1) Conceptual: it makes formal and explicit \textbf{why} the lower bound holds, namely due to the transfer of the lower bound for the problem without migrations to the problem with migrations. 2) Generality: Theorem  \ref{thm:subInformal}, without modifications in its proof, can be used to show lower bounds for other problems. In particular, it leads to the following interesting consequence: For the clairvoyant version of \dyn (where the items' durations are revealed upon arrival), the proof of Theorem \ref{thm:subInformal} still applies and shows that sublinear migrations do not help randomized algorithms. This is despite the fact that no lower bound for randomized algorithms without migrations is currently known! (\cite{AzarVainstein} gives a lower bound for deterministic algorithms).

\medskip 
\noindent \textbf{Using a linear amount of migrations (\Cref{sec:migration-costs}).} 
Suppose our goal is a $\frac{1}{\alpha}$-approximation for some $\alpha < 1$. A typical approach to achieve this is to ensure that almost all open bins have load at least $\alpha$. To maintain this property, we need migrations to handle bins with small remaining load after some departures. However, it is not enough to migrate away from a bin when its load drops below $\alpha$. To see this, consider the following example. It could be the case that the bin's load was only a single item above load $\alpha$, so a single departure triggers a large amount of migrations away from this bin. Thus, while such a migration rule would maintain a $\frac{1}{\alpha}$-approximation, we can do much more than $\alpha n$ migrations in the worst case.

To fix this issue, we introduce a second condition in the migration rule: a bin is only eligible for migration if its load drops below $\alpha$ \emph{and it was previously ``almost full'' (i.e. its load was above some particular threshold $\approx 1$)}. With additional arguments to handle different item sizes, this modification gives the desired $\frac{1}{\alpha}$-approximation (up to additive terms) with $\approx \alpha n$ migrations, because we can charge the number of migrations to the previous almost full load of the bin.

\medskip 
\noindent \textbf{Dynamic Bin Packing with delays  (\Cref{sec:migration-delays}).} 
Going back to \Cref{fig:bad_example}, we can see the reason we lose a $\mu$-factor is because we open almost all bins for a $\mu$-factor longer than they ``needed to be.'' This is unavoidable if we are non-clairvoyant, and cannot migrate, but what if we can do migrations subject to migration delays?

Concretely, consider the same instance: the bad allocation is unavoidable at time $0$ due to non-clairvoyance, but at time $1$ we can see that each open bin now has a single active item after all the short-lived ones depart. Suppose $C \geq 1$. We do not know the remaining duration of these items, but they can all fit in a single bin together. When is it worth it to migrate them? At one extreme, if we migrate them now at time $1$, it could be the case that their duration is $\approx 1$, so the migration delay causes us to open a single bin for a $C$-factor longer than it needed to be. 
At the other extreme, suppose we wait until time $C$; now, we have a certificate that these items have duration $> C$, so the migration delays only cost us a constant factor. However, now the bins have already been open for a $C$-factor longer than they needed to be (since we could have fixed this packing at time $1$ if we had known). A moment's thought shows that the right migration schedule is to perform a migration after $\sqrt{C}$ units of \emph{real processing} (not counting the migration delays) is done. This guarantees that the items' durations are not blown-up by more than a $\sqrt{C}$-factor.

However, this is not enough: we need to migrate the items to the right places. For that, we employ a known $O(\mu)$-competitive online algorithm in a way that the ``effective'' $\mu$ is $O(1)$ (ignoring the beginning of the items, which are treated separately); this gives that our algorithm is $O(1)$-competitive against the optimal solution of the problem with the blown-up durations. The final step is to show that blowing up the items' durations by at most a $\sqrt{C}$-factor, does not increase $\OPT$ by more than $O(\sqrt{C})$.


\subsection{Additional Related Work}

The Bin Packing problem and its many variations have been intensely studied since the 50s. Since the literature is too vast to cover here, we focus only on the directions that help contextualizing our work and refer to~\cite{BinPackingSurvey} for additional references. 

\medskip 
\noindent \textbf{Offline and non-dynamic Online Bin Packing.} Even in the offline case where all items are known upfront, the Bin Packing problem is NP-Hard but can be approximated within an additive term of $O(\log \OPT)$ \cite{rothvoss}. In the online but non-dynamic version of the problem, i.e., items do not depart the system, Balogh et al.~\cite{binPackingWorstCaseESA} gave a 1.5783-competitive algorithm, and a lower bound of 1.5404 was given by~\cite{baloghLB}. This non-dynamic version has also been studied in the stochastic arrival model, see for example~\cite{rhee1993line,csirik2006sum,binPackingVarun,binPackingKnownT}.

\medskip

\noindent \textbf{\dyn without migrations.} We move on to the dynamic version of the problem, \dyn, where items arrive and depart over time \cite{coffman1983dynamic}. The \dyn with the total active (usage) time objective \cite{nonClairvoyantDynBP, li2015dynamic} has been studied in both the non-clairvoyant and the clairvoyant settings. In the non-clairvoyant case, \cite{tang2016first} showed that the First-Fit algorithm has a competitive ratio of $\mu+4$, where $\mu$ is the ratio of the longest to shortest item duration. This is almost optimal as \cite{li2015dynamic} showed a lower bound of $\mu+1$ on the competitive ratio of any fit-packing algorithm. 
The clairvoyant case, where the duration of an item is revealed upon its arrival, was introduced more recently by \cite{ren2016clairvoyant}. The authors provided a $O(\frac{\log \mu}{\log \log \mu})$-competitive algorithm, and showed a constant lower bound of $\frac{1+\sqrt{5}}{2}$ on the competitive ratio of any online packing algorithm. Azar and Vainstein~\cite{AzarVainstein} improved upon this result by presenting a $O(\sqrt{\log \mu})$-competitive algorithm in this setting, and also showed that this is best possible. The papers~\cite{DBPpredictions,DBPKonstantina}  interpolate between the non-clairvoyant and clairvoyant models by assuming that predictions for the items' durations (and/or future load) are available, obtaining guarantees that depend on the quality of the predictions. 









\medskip 
\noindent \textbf{\dyn with migrations.}
The tight upper and lower bounds on the competitive ratio of \dyn using $>n$ migrations are given in \cite{feldkord2018fully}, for the stronger objective of being competitive in each time step: for any $0 < \epsilon < 1$, they obtain a $\approx 1.387 + \epsilon$-competitive algorithm with $O(\frac{1}{\epsilon})$ additive term and at most $O(\frac{n}{\epsilon^2})$ migrations. In the size-cost model of migrations, they obtain a $(1+\e)$-competitive algorithm with $O(\frac{1}{\epsilon^2})$ additive term where  the total size of migrated items is at most $O(\frac{1}{\e})$ times the total size of the instance, also for the stronger per-time competitive objective. Berndt et al. \cite{berndt} obtain a similar result (with improved additive term $O(\frac{1}{\e} \log \frac{1}{\e})$) with a stricter limit on the migrations: the total size of migrations performed at a given time is at most $O(\frac{1}{\e^4} \log \frac{1}{\e})$ times the size of the current item. They also prove that these results are essentially optimal~\cite{berndt,feldkord2018fully}. Finally, \cite{feldkord2018fully} also consider a more general model of migration cost, where each item has its own non-negative cost to migrate it. They then leverage the $\approx 1.589$-competitive algorithm for the static Bin Packing problem from \cite{seiden} to obtain essentially the same approximation in the fully-dynamic case while incurring migration cost that is at most a constant times the total cost of all items.


\medskip 
\noindent \textbf{Other objectives.} Another popular objective function for the Dynamic Bin Packing problem is that of minimizing the maximum number of bins used, that is, to obtain the minimum value $k$ such that all items in the instance can be packing within $k$ bins. This objective measures the cost of a solution in a coarser way that does not track for how long these bins are actually used. Under this objective, \cite{coffmanGJ} designed an algorithm that is 2.897-competitive without using any migration (note that such strong guarantees are not possible for our objective), and~\cite{DBPMaxLB} showed a lower bound of $8/3 \approx 2.66$ in this setting. 

\medskip

\noindent \textbf{Other optimization problems with recourse.} A variety of other combinatorial optimization problems have been studied in the dynamic setting where objects arrive and depart over time, and the algorithm is allowed some \emph{recourse} to modify past decisions in light of these changes. The goal is to maintain a good solution in terms of objective value with limited recourse (in terms of number of changes or running time of those changes). These problems include load balancing/machine scheduling \cite{DBLP:journals/mor/SandersSS09, DBLP:conf/innovations/0004KSS23, DBLP:conf/stoc/Krishnaswamy0S23,albersMig}, discrepancy minimization/vector balancing \cite{DBLP:conf/soda/GuptaGKKS22}, steiner tree \cite{DBLP:journals/siamdm/ImaseW91, DBLP:conf/soda/GuptaK14, DBLP:conf/stoc/LackiOPSZ15}, facility location/clustering \cite{DBLP:conf/nips/Cohen-AddadHPSS19,DBLP:conf/approx/GuoKLX20, DBLP:conf/aistats/GuoKLX21}, and set cover \cite{DBLP:conf/stoc/GuptaK0P17, DBLP:conf/focs/GuptaL20, DBLP:journals/siamcomp/BhattacharyaHNW23}.

\section{Sublinear Migrations Do Not Help} \label{sec:sublinear}

    In this section we prove our first claimed result, namely that for \dyn a sublinear amount of migrations does not buy significant additional power compared to not using any migration. To state the result precisely, let $\cI_\mu$ be the family of all instances of \dyn where the items have duration between $1$ and $\mu$ (recall that in general, $\mu$ is defined to be the ratio of the longest to shortest item duration).
    We use $\cost(\cA, I)$ to denote the cost (total active time) of the algorithm $\cA$ on the instance $I$, and $\mig(\cA, I)$ the number of migrations it makes on this instance. For a randomized algorithm $\cA_R$, we use $\E_R$ to denote expectation with respect to the algorithm's randomness.

		

\begin{restatable}{thm}{thmSublinear}\label{thm:sublinear}
		Consider a value $\mu > 1$, and $\beta = \beta(\mu)$. Suppose there is a finite subfamily of instances $\cI'_\mu$ of $\cI_\mu$ that is $\beta$-hard for every randomized algorithm $\cA_R$ that uses zero migrations, i.e., there is $I \in \cI'_\mu$ such that $\E_R\,\cost(\cA_R, I) \ge \beta \OPT(I)$.
		
		Consider any randomized algorithm $\cA_R$ that makes a sublinear expected number of migrations on the instances $\cI_{\mu}$, i.e., $\max_{I \in \cI_\mu : |I| = n} \E_R\,\mig(\cA_R, I) \in o(n)$. Then the algorithm $\cA_R$ is no better than $\frac{\beta}{3}$-competitive, namely there is an instance $I \in \cI_\mu$ such that $\E_R\,cost(\cA_R, I) \ge \frac{\beta}{3} \OPT(I)$.
\end{restatable}

    Employing this transfer result to the lower bound of $\Omega(\mu)$ for algorithms for \dyn that make zero migrations~\cite{nonClairvoyantDynBP},\footnote{\cite{nonClairvoyantDynBP} only states the lower bound for deterministic algorithms, but as we show in Appendix \ref{app:basicLB} it holds for randomized algorithms as well. Further, the hard family of instances that we construct is finite.} we obtain the following consequence.  

	\begin{cor}\label{cor:sublinear}
		Every randomized algorithm for \dyn that makes $o(n)$ migrations in expectation has competitive ratio at least $\frac{\mu}{6}$.
	\end{cor}
	
	Since the competitive ratio of $O(\mu)$ can be obtained without any migrations~\cite{tang2016first}, we conclude that a sublinear number of migrations cannot improve the competitive ratio by more than a constant factor. 
	
	\medskip
	
	For the remainder of the section we prove \Cref{thm:sublinear}. As mentioned in the introduction, the strategy is to show the contrapositive, namely that given a randomized algorithm $\cA_R$ with sublinear migrations that is $\beta$-competitive, we can extract an algorithm $\cA'_R$ with no migrations that is $\Theta(\beta)$ competitive; the hardness of the latter would then translate to the same hardness (up to constant factors) to the former. The desired extraction can be equivalently stated as \eqref{eq:mainLB}. 
 
    Also as mentioned, we cannot show this directly. Then first we show the following weaker version of \eqref{eq:mainLB} where the quantifiers in ``there is an algorithm, such that for every (distribution over) instances'' are exchanged, and we get a ratio-of-expectations instead of expectation-of-ratio.     

	\begin{lemma} \label{lemma:tensor}
		Consider $\mu > 0$ and a finite family of instance $\cI'_{\mu} \subseteq \cI_\mu$. For some $\beta = \beta(\mu)$, suppose there is a randomized online algorithm $\cA_R$ that is $\beta$-competitive for the instances in $\cI'_\mu$, and that makes a sublinear expected number of migrations on these instances. Then for any distribution $\nu$ over $\cI'_\mu$, there is a randomized algorithm $\cA'_R$ that makes 0 migrations and has expected cost $\E_{I \sim \nu, R}\, \cost(\cA'_R, I) \le 3 \beta \cdot \E_{I \sim \nu}\, \OPT(I)$.
	\end{lemma}
	
	\begin{proof}
            Fix a distribution $\nu$ over $\cI'_\mu$. Let $n$ be the largest number of items on an instance in the support of $\nu$. Consider a small enough $\e > 0$ (letting $\e = \frac{\beta}{2n \mu} \cdot\E_{I \sim \nu}\,\OPT(I)$ suffices). Let $f(m) := \max_{I \in \cI'_\mu : |I| = m} \E_R\, \mig(\cA_R, I)$ denote the worst expected number of migrations for instances of size $m$. By assumption $f$ is sublinear, which implies that $\lim_{k \rightarrow\infty} \frac{f(kn)}{k} = 0$, since this limit equals $n \cdot \lim_{k \rightarrow\infty} \frac{f(kn)}{nk} = n \cdot 0$. Then let $\bar{k}$ be large enough so that $\frac{f(\bar{k}n)}{\bar{k}}  \le \e$. 
	
		Sample $\bar{k}$ instances $I_1,\ldots,I_{\bar{k}}$ i.i.d. from $\nu$, and let $I_{\full} := I_1 
  \oplus \ldots \oplus I_{\bar{k}}$ be the concatenation of these instances (throughout, we always concatenate instances without overlap between items coming from different instances). Run $\cA_R$ over this full instance $I_{\full}$. For $j = 1,\ldots,\bar{k}$, let $\cost_j$ and $\mig_j$ be respectively the cost and number of migrations that this algorithm does on the part of the instance coming from $I_j$; notice these are random variables that depend only on the prefix instance $I_1 \oplus \ldots \oplus I_j$ and on the randomness of the algorithm. 
		
		The first claim is that there is an index $\bar{j} \in [\bar{k}]$ such that $\E\, \cost_{\bar{j}} \le 2\beta \cdot \E\, \OPT(I_0)$ (where $I_0 \sim \nu$) and $\E\, \mig_{\bar{j}} \le 2\, \frac{f(\bar{k}n)}{\bar{k}} \le 2\e$ (the last inequality by the definition of $\bar{k}$). Since the instances that constitute $I_\full$ do not overlap, the cost of the algorithm $\cA_R$ on $I_\full$ equals $\sum_j \cost_j$, and likewise the optimal cost for the full instance can be decomposed as $\OPT(I_{\full}) = \sum_j \OPT(I_j)$. Since $\cA_R$ is $\beta$-competitive (in particular for $I_{\full}$), this gives 
	\begin{gather*}
		\sum_{j = 1}^{\bar{k}} \E\, \cost_j \le \beta \sum_{j = 1}^{\bar{k}} \E\, \OPT(I_j)  = \beta \bar{k} \cdot \E\, \OPT(I_0), 
	\end{gather*}
	where the last equation is because $I_0$ and the $I_j$'s have the same distribution, $\nu$. Also, the total expected number of migrations that  $\cA_R$ makes on $I_{\full}$ is $\sum_{j=1}^{\bar{k}} \E\, \mig_j \le f(\bar{k} n)$. Thus, we have the averages
		\begin{gather*}
			\frac{1}{\bar{k}} \sum_{j = 1}^{\bar{k}} \E\, \cost_j \le \beta\, \E\, \OPT(I_0) \textrm{~~~and~~~} \frac{1}{\bar{k}} \sum_{j=1}^{\bar{k}} \E\, \mig_j \le \frac{f(\bar{k}n)}{\bar{k}}.
		\end{gather*}		
    Since the quantities $\E\, \cost_j$ and $\E\, \mig_j$ are non-negative, applying Markov's inequality to the averages we see that fewer than half of the $j$'s have $\E\, \cost_j > 2 \beta \cdot \E\, \OPT(I_0)$ and fewer than half of the $j$'s have $\E\, \mig_j > 2\,\frac{f(\bar{k}n)}{\bar{k}}$. Thus, there is a $\bar{j}$ where we have both $\E\, \cost_{\bar{j}} \le 2 \beta \cdot \E\, \OPT(I_0)$ and $\E\, \mig_{\bar{j}} \le 2 \frac{f(\bar{k}n)}{\bar{k}} \le 2\e$, as claimed. 
		
		Now we define the intermediate algorithm $\tilde{\cA}_R$ any $n$-job instance $I$ as follows:  We sample ``simulated'' instances $\tilde{I}_1,\ldots,\tilde{I}_{\bar{j}-1}$ from $\nu$, run $\cA_R$ over the concatenated instance $\tilde{I}_1 \oplus \ldots \oplus \tilde{I}_{\bar{j}-1} \oplus I$, and only use the decisions that the algorithm makes on the part $I$ of the instance. 
	       
    Consider $I \sim \nu$, so the instances $\tilde{I}_1 \oplus \ldots \oplus \tilde{I}_{\bar{j}-1} \oplus I$ and $I_1 \oplus \ldots \oplus I_{\bar{j}-1} \oplus I_{\bar{j}}$ have the same distribution. By construction, the expected cost of $\tilde{\cA}_R$ on the instance $I$ equals the expected cost that $\cA_R$ incurs in the part $I$ of the instance $\tilde{I}_1 \oplus \ldots \oplus \tilde{I}_{\bar{j}-1} \oplus I$, which is the expected cost that it incurs on the $I_{\bar{j}}$ part of the instance $I_1 \oplus \ldots \oplus I_{\bar{j}-1} \oplus I_{\bar{j}}$, which is exactly the 
cost $\E\, \cost_{\bar{j}}$ above. Thus, $\E\,\cost(\tilde{\cA}_R, I) \le 2 \beta \cdot \E\,\OPT(I_0) = 2 \beta \cdot \E\,\OPT(I)$. For the same reason, $\E\,\mig(\tilde{\cA}_R, I) = \E\, \mig_{\bar{j}} \le 2 \e$. By Markov's inequality, we further get that the probability that $\tilde{\cA}_R$ makes any migrations on $I$ is $\Pr(\mig(\tilde{\cA}_R, I) \ge  1) \le 2 \e$. 

	The final algorithm $\cA'_R$ that makes no migrations is obtained as follows: we run $\tilde{\cA}_R$, and if it makes a migration we just ignore it (and all subsequent migrations) and place each future item in its own bin. 
	
	To conclude the proof we just need to bound the expected cost of the algorithm $\cA'_R$, namely to show that for an instance $I \sim \nu$ we have $\E\,\cost(\cA'_R, I) \le 3 \beta \cdot \E\,\OPT(I)$. To obtain this bound, let $B$ be the event that $\tilde{\cA}_R$ makes a migration when running on instance $I$, which happens with probability at most $2 \e$. Under the complement event $B^c$, the algorithms $\tilde{\cA}_R$ and $\cA'_R$ make the same decisions on $I$ and so $\cost(\cA'_R, I) = \cost(\tilde{\cA}_R, I)$, and under the event $B$ we have $\cost(\cA'_R, I) \le n \mu$, since $I$ has $n$ jobs of duration at most $\mu$. Therefore,
        \begin{align*}
            \E\,\cost(\cA'_R, I) \,\le\, \E\,\cost(\tilde{\cA}_R, I) + 2 \e n \mu \,\le\, 3 \beta \cdot \E\,\OPT(I),
        \end{align*}
        where the last inequality uses the definition of $\e$. This concludes the proof. 
        \end{proof}

	Next, we prove a minimax theorem that ``boosts'' the previous lemma and shows that it actually implies the stronger result \eqref{eq:mainLB}. We state the theorem in more generality, to make it convenient for its possible future use in different contexts. Here, $\cX$ should be thought as the family of all deterministic algorithms, $\Y$ the set of all instances to a problem, $g(x,y)$ the cost of algorithm $x$ on instance $y$, and $h(y)$ as the optimum for the instance $y$. With this interpretation, the ``$\star$'' equality in the theorem implies the following: if for every distribution $\nu$ of instances there is a randomized algorithm $\cA'_R$ with $\frac{\E_{I \sim \nu, R} \cost(\cA'_R, I)}{\E_{I \sim\nu} \OPT(I)} \le \Theta(\beta)$ (which is that \Cref{lemma:tensor} gives), then there is a randomized algorithm $\cA''_R$ where for all instances $I$ we have $\frac{\E_R \cost(\cA''_R, I)}{\OPT(I)} \le \Theta(\beta)$, which is equivalent to the desired property~\eqref{eq:mainLB}. In the statement, we use $\Delta(U)$ to denote the set of all probability distributions over a given finite set $U$.

	\begin{thm} \label{thm:extendedMinimax}
		Let $\cX$ and $\Y$ be finite sets, $g : \cX \times \Y \rightarrow \R_{>0}$ be a strictly positive function, and $h : \Y \rightarrow \R_{\ge 0}$ be a non-negative function. Then (in all terms $X \sim p$ and $Y \sim q$)
		\begin{align*}
			\max_{q \in \Delta(\Y)} \min_{x \in \cX} \Es \frac{g(x,Y)}{h(Y)} = \min_{p \in \Delta(\cX)} \max_{y \in \Y} \frac{\Es g(X,y)}{h(y)} \stackrel{\star}{=} \max_{q \in \Delta(\Y)} \min_{p \in \Delta(\cX)} \frac{\Es g(X,Y)}{\Es h(Y)} = \min_{p \in \Delta(\cX)} \max_{q \in \Delta(\Y)} \frac{\Es g(X,Y)}{\Es h(Y)}. 
		\end{align*}
	\end{thm}

	\begin{proof}
		Since the function $(p,q) \mapsto \E_{X \sim p, Y \sim q}\, \frac{g(X,Y)}{h(Y)}$ is bilinear in $p$ and $q$, the first equation follows from the standard von Neumann's minimax theorem~\cite{stdMinimax}. The last equality, which is the hardest one, follows from the minimax Theorem 2.14 of~\cite{DBLP:journals/corr/abs-2002-10802}, since the functions $(p,q) \mapsto \E_{X \sim p, Y \sim q}\, g(X,Y)$ and $(p,q) \mapsto \E_{Y \sim q}\, h(Y)$ are bilinear (and $g$ is strictly positive).
		
		Given this, to prove equation ``$\star$'' it suffices to it suffices to show
		\begin{align}
			\min_{p \in \Delta(\cX)} \max_{y \in \Y} \frac{\Es g(X,y)}{h(y)} \ge \max_{q \in \Delta(\Y)} \min_{p \in \Delta(\cX)} \frac{\Es g(X,Y)}{\Es h(Y)} \textrm{~~and~~} \min_{p \in \Delta(\cX)} \max_{q \in \Delta(\Y)} \frac{\Es g(X,Y)}{\Es h(Y)} \ge \min_{p \in \Delta(\cX)} \max_{y \in \Y} \frac{\Es g(X,y)}{h(y)}. \label{eq:minimaxRatio}
		\end{align}
		To show the first of these, let $L := \min_{p \in \Delta(\cX)} \max_{y \in \Y} \frac{\E_{X \sim p}\, g(X,y)}{h(y)}$ denote its left-hand side, and let $\bar{p}$ be the distribution attaining this $\min$. This means that for all $y \in \Y$, $\E_{X \sim \bar{p}}\, g(X,y) \le L \cdot h(y)$, and so by averaging, for any distribution $q \in \Delta(\Y)$ we have $\E_{Y \sim q} \E_{X \sim \bar{p}}\, g(X,Y) \le L \cdot \E_{Y \sim q} h(Y)$. Thus,
		\begin{align*}
			\max_{q \in \Delta(\Y)} \min_{p \in \Delta(\cX)} \frac{\E_{X \sim p, Y \sim q}\, g(X,Y)}{\E_{Y \sim q}\, h(Y)} \le \max_{q \in \Delta(\Y)} \frac{\E_{Y \sim q} E_{X \sim \bar{p}}\, g(X,Y)}{\E_{Y \sim q}\, h(Y)} \le L,
		\end{align*}
		the last inequality following from the previous observation. This proves the first inequality in \eqref{eq:minimaxRatio}.
		
		Finally, the second inequality in \eqref{eq:minimaxRatio} is true just because the distribution that puts all the mass on a single $y$ is a feasible option for $q$. This concludes the proof. 
	\end{proof}

Having \Cref{lemma:tensor} and \Cref{thm:extendedMinimax}, we can now prove the main result of this section. 
	
	\begin{proof}[Proof of \Cref{thm:sublinear}]
		Let $\cI'_\mu$ be the finite family of instances for \dyn from the statement of the theorem, namely where every randomized online algorithm with zero migrations is at least $\beta$-competitive. Let $NoMig$ denote the set of all deterministic online algorithms for \dyn  that use no migrations. Since $\cI'_\mu$ is finite, by identifying the algorithms that have the same behavior over all instances in $\cI'_\mu$ we can assume without loss of generality that $NoMig$ is also finite.
						
		Now suppose for contradiction that there is a randomized algorithm $\cA_R$ that makes a sublinear expected number of migrations on the instances $\cI_\mu$ and is strictly better than $\frac{\beta}{3}$-competitive for these instances (so in particular for the subset of instances $\cI'_\mu \subseteq \cI_\mu$). Employing \Cref{lemma:tensor}, this means that for every distribution $\nu$ over $\cI'_\mu$, there is a randomized algorithm $\cA'_R$ with zero migrations and $\E_{I \sim \nu, R}\, \cost(\cA'_R, I) < \beta \cdot \E_{I \sim \nu} \OPT(I)$; thus, by identifying randomized algorithms with distributions over deterministic algorithms, we have $$\max_{\nu \in \Delta(\cI'_\mu)} \min_{p \in \Delta(NoMig)} \frac{\E_{I \sim \nu, \cA \sim p}\,\cost(\cA, I)}{\E_{I \sim \nu} \OPT(I)} < \beta.$$ But then employing \Cref{thm:extendedMinimax}, its equation ``$\star$'' guarantees the existence of a distribution $p \in \Delta(NoMig)$ with $\max_{I \in \cI'_{\mu}} \frac{\E_{\cA \sim p}\,\cost(\cA,I)}{\OPT(I)} < \beta$, i.e., a randomized algorithm that makes no migrations and has competitive ratio strictly better than $\beta$ over the family of instance $\cI'_\mu$. This contradicts the $\beta$-hardness of the family $\cI'_\mu$, thus concluding the proof. 
	\end{proof}


\section{$<n$ Migrations Do Help} \label{sec:migration-costs}

Given the negative result in the previous section, we cannot hope for improved algorithms using $o(n)$ migrations. Thus, here we zoom in on using a linear number $\alpha n$ of migrations, where we can use $\alpha$ as a finer control on the amount of migrations, and explore how the competitive ratio depends on $\alpha$. We prove the following tradeoff (stated in the introduction, reproduced here for convenience) for the stronger objective of keeping a competitive packing on a per-time basis. Recall that $\OPT_t$ denotes the minimum number of bins necessary for packing the items that are in the system at time $t$, so we are comparing our algorithm to the optimal policy that can arbitrarily re-pack items at any time.


\unitMigrationCosts*


As mentioned in the introduction, in \Cref{lem:migrationtradeoff} we show that this tradeoff between the number of migrations and competitive ratio is almost tight, namely that every randomized algorithm that does at most $\alpha n$ migrations has competitive ratio at least $\Omega(\frac{1}{\alpha})$. \blue{We defer the proof of this lower bound to \Cref{app:unit-migration-LB}, and for the remainder of the section we define the algorithm guaranteed by \Cref{thm:unit-migration-costs} and prove that it achieves the desired upper bound.}








\subsection{Algorithm}

The key idea in our algorithm is to classify bins into two types (\emph{Bad}, and \emph{Good}) depending on how their loads have evolved. Each bin is initially labeled as Bad when it gets assigned its first item. The algorithm can then keep assigning items to the bin. A bin transitions from Bad to Good if at some point it becomes almost full (determined by a parameter $f > \alpha$). For a Good bin, if departures cause its load to drop below a threshold (determined by the parameter $\alpha$), its remaining load gets migrated to different bins; this migration is performed using \firstfit, which maintains an ordering of the open bins, and assigns each arriving item to the earliest bin in the order where it fits (or opens a new bin if needed). This whole procedure is detailed in  \Cref{AlgMigrationCosts}. 

\begin{algorithm}
	\SetAlgoLined
	\DontPrintSemicolon
	Upon arrival of a new item $i$ with size $s_i$: \;
	\Begin{If there is a Bad bin with load $\leq 1-s_i$, assign $i$ to this bin. If its load becomes $\geq f$, label the bin as Good.\;
		Otherwise, if there are Good bins with load $\leq 1-s_i$, assign $i$ to any of these bins. \;
		Otherwise, open a new bin, assign $i$ to it, and label it as Bad if its load is $< f$ and Good otherwise. \;}
	
	Upon departure of an item $i$: \;
	\Begin{
		If $i$'s bin was Good and now is left with load $< \alpha$, migrate all of its remaining jobs using \firstfit with the bins ordered as follows: Bad bins, then Good bins, then new bins.
	}
	\caption{Single-class algorithm with bounded migration (parameters $\alpha$, $f$)} \label{AlgMigrationCosts}
\end{algorithm}

The idea why \Cref{AlgMigrationCosts} works is the following: Since we do not allow Good bins to fall below load $\alpha$, a volume argument shows that the number of such bins is at most $\frac{1}{\alpha}$ times $\OPT_t$, in every time $t$. Also, since Good bins never become Bad, by choosing $f$ appropriately we should also be able to control the number of Bad bins at any time; together, these will determine the competitive ratio of our algorithm. Moreover, observe that a bin must become almost full (load $\geq f$) before being migrated. This should allow us to charge the migrations away from this bin to the departures of the items that initially made the bin Good; roughly speaking, we migrate out at most an $\alpha$ volume for each $\ge f$ (think $f \approx 1$) volume of insertions, upper bounding the migrations.


The issue with this analysis is that the last part of the argument does not bound the \emph{number} of migrations performed. In order to relate number of items and volume, we partition the items into classes based on their sizes, in powers-of-two intervals, and employ  \Cref{AlgMigrationCosts} on each class separately. More precisely, we say that an item with size in $(\frac{1}{2^{c+1}}, \frac{1}{2^c}]$ has \emph{class $c$}, for $c=0,1,\ldots$. Elements of class more than $\log \act - 1$, i.e., with size at most $\frac{1}{\act}$, are considered ``small'' and treated separately, since we are guaranteed that a single ``junk bin'' suffices to accommodate them. Unfortunately the maximum number of simultaneous items in the system $\rho$ is not known upfront, so our final algorithm uses a guess-and-double approach to estimate it. This final algorithm is described in \Cref{AlgMigrationCostsClass}. 



\begin{algorithm}
	\SetAlgoLined
	\DontPrintSemicolon
	Initialize guess $\tilde{\act} = 1$.\;
	Initialize an instance of \Cref{AlgMigrationCosts} for items of size class $0$ with parameters $(\alpha, f_0)$, where $f_0 = \frac{1}{2}$  \;
	Open a single \emph{junk bin} for the current guess of $\tilde{\act}$.\;
        \For{each arriving item}{
		\If{the current number of items in the system is $\ge \tilde{\act}$}{
			Double the guess $\tilde{\act}$.\;
			Initialize a new instance of \Cref{AlgMigrationCosts} for items of size class $c = \log \tilde{\act}$ with parameters $(\alpha, f_c)$, where $f_c = 1- \frac{1}{2^c}$. \;   
			Open a new junk bin for this guess $\tilde{\rho}$.\;			
		}
		\If{the incoming item belongs to a class $c < \log \tilde{\rho}$}{Assign it to a bin using the instance of \Cref{AlgMigrationCosts} for class $c$.\;}
		\Else{
			Assign it to the junk bin for the current guess $\tilde{\rho}$. \label{alg:junk}\;
		}
		
	}
	\caption{Complete algorithm with bounded migration (parameter $\alpha$)} \label{AlgMigrationCostsClass}
\end{algorithm}

 \subsection{Analysis} \label{sec:migCostAnalysis}

 It is convenient to partition the times into \emph{phases} corresponding to distinct guesses of $\tilde{\act}$. Note that by the doubling procedure, the guess $\tilde{\act}$ is always at most $2 \act$; in particular, there are at most $\log \act + 1$ phases. 
Also, notice that for a particular guess $\tilde{\rho}$, items that have size class $\ge \log \tilde{\rho}$ (which are assigned to the junk bin) have size at most $\frac{1}{\tilde{\rho}}$; we call such items \emph{small}. 

We start by verifying that Line \ref{alg:junk} in \Cref{AlgMigrationCostsClass} is well-defined, that is, all small items in a phase can indeed fit in the single junk bin of the current guess. 

\begin{lemma}\label{lemma:junk}
	For each phase, every small item fits into the junk bin of the current guess upon arrival.
\end{lemma}

\begin{proof}
During a phase $\tilde{\act}$, the number of items currently in the system is always at most  $\tilde{\act}$, The total size of small items currently in the system is then at most $\tilde{\act} \cdot \frac{1}{\tilde{\act}} = 1$, and so they all fit in the single junk bin.    
\end{proof}

We now bound the number of bins opened at any time by our algorithm. 



\begin{lemma}\label{lem:per-time}
  At every time $t$, 
 \Cref{AlgMigrationCostsClass} has at most $\frac{1}{\alpha} \OPT_t + O(\log \act)$ bins open.
\end{lemma}
\begin{proof}
    First, we open only one junk bin per phase, thus we open at most $O(\log \act)$ junk bins at any point in time. To bound the number of Bad bins, we have the following claim. 

	
	\begin{claim}\label{claim:num-bad}
		 Each instance of \Cref{AlgMigrationCosts} maintained by \Cref{AlgMigrationCostsClass} has at most one Bad bin at each time.
	\end{claim}
	\begin{proof}
     Consider the instance of \Cref{AlgMigrationCosts} for a size class $c$, which then receives the items with sizes in $(\frac{1}{2^{c+1}}, \frac{1}{2^c}]$. For class $c=0$, the parameter $f$ in this instance was set to $f_0 = \frac{1}{2}$. Since every item in class 0 has size strictly larger than $\frac{1}{2}$, as soon as we place such item in a bin it becomes Good. Thus, there is no Bad bin for this class. 

     Now consider a class $c > 0$, and recall that for this instance of  \Cref{AlgMigrationCosts} we set the parameter $f$ to be $f_c = 1-\frac{1}{2^{c}}$. Assume by contradiction that at some point a second Bad bin is opened by this instance. That is, there currently exists a Bad bin with load $\ell < f_c$, but the arrival or migration of an item with size $s \in (\frac{1}{2^{c+1}}, \frac{1}{2^c}]$ causes another Bad bin to be opened. Since this item was not placed in the existing Bad bin, it must not have fit, namely $\ell + s > 1$. Combining these inequalities gives
		$1 < \ell + s < (1 - \frac{1}{2^c})  + \frac{1}{2^c}$, which is a contradiction.
	\end{proof} 

    Since each phase starts only one new instance of \Cref{AlgMigrationCosts}, we also have at most $O(\log \act)$ Bad bins at any point in time. 
    
    It remains to bound the contribution of the Good bins at a given time $t$. By definition of \Cref{AlgMigrationCosts}, every Good bin has load at least $\alpha$, since when its load drops below it the remaining items are migrated out of it and the bin is closed. Thus we have:
	\[(\text{number of Good bins at time $t$}) \leq \frac{1}{\alpha}(\text{total size of items at time $t$}) \leq \frac{1}{\alpha} \OPT_t.\]
	Summing the contributions from the junk, Bad, and Good bins completes the proof of the lemma. 
\end{proof}


Finally, we bound the number of migrations that  \Cref{AlgMigrationCostsClass} performs over the whole horizon.


\begin{lemma}\label{lem:num-migration}
	\Cref{AlgMigrationCostsClass} makes at most $\frac{4 \alpha}{1 - 2 \alpha} \cdot n$ migrations.
\end{lemma}

\begin{proof}
    It suffices to show that for each $c = 0,1,\ldots$, the instance of \Cref{AlgMigrationCosts} for the class size $c$ does at most $\frac{4 \alpha}{1 - 2 \alpha} \cdot n_c$ migrations, where $n_c$ is the number of items in that class. Since migrations are only done within \Cref{AlgMigrationCosts}, summing over all size classes would complete the proof.
   
	
	Thus, consider any size class $c$. \Cref{AlgMigrationCosts} only migrates when a Good bin's load decreases below $\alpha$, at which time we migrate all items away from this bin and close it. Prior to this event, the bin must have become Good at some point, meaning its load was at least $f_c$. Between these two events, it must be the case that at least a $(f_c - \alpha)$-volume of items depart from this bin (not via migration, but rather leaving the system). Every item in this size class has size in $(\frac{1}{2^{c+1}}, \frac{1}{2^c}]$, so this $(f_c - \alpha)$-volume of items corresponds to at least $2^c(f_c - \alpha)$ items departing. From this bin we migrate at most an $\alpha$-volume of items (\Cref{AlgMigrationCosts} only migrates from Good bins of current volume less than $\alpha$), which corresponds to at most $2^{c+1} \alpha$ migrations. Thus for each bin, we have:
	\[(\text{number of migrations away from this bin}) \leq \frac{2^{c+1} \alpha}{2^c(f_c - \alpha)} (\text{number of departures from this bin}).\]
	Summing over all bins in the instance of this class $c$ gives that the number of migrations in this size class is at most $\frac{2^{c+1} \alpha}{2^c(f_c - \alpha)} n_c \leq \frac{2 \alpha}{\frac{1}{2} - \alpha} n_c = \frac{4 \alpha}{1 - 2 \alpha} n_c$, as required.
\end{proof}

Together, Lemmas \ref{lem:per-time} and \ref{lem:num-migration} prove \Cref{thm:unit-migration-costs}.

\section{Dynamic Bin Packing with Migration Delays } \label{sec:migration-delays}


Here we introduce the problem \emph{fully dynamic bin packing with delays} (\dynd): instead of migrations incurring some global cost (i.e. unit- or size-costs) as in the standard \dyn, now every time an item $i$ is migrated its duration is increased by $+ C$, where the delay cost $C$ is an input parameter known to the algorithm. As before, we compare to the optimal policy that can arbitrarily re-pack at any time -- further, the optimal policy does not incur any delay cost for re-packing.

More precisely, as before, an instance of dynamic bin packing arrives online (items arrive at their release times, revealing their size, but not their duration). If the algorithm migrates an item, its duration is increased by an additive $C$ (known upfront), which we call the \emph{delay cost}. The goal is to minimize the total active time $\sum_{i} \int_{t = 0}^\infty \ones(\text{bin $i$ open at time $t$})\,\d t$. We call this problem \dynd. \blue{We compare our algorithm to the same optimal offline policy as in \dyn; in particular, it knows all items upfront and can repack arbitrarily (with no delay cost).} Notice that unlike in the previous section, we do not need to be competitive in each time step; such stronger metric is not well-defined for \dynd, since there may be a time where $\OPT$ has no items in the system but the algorithm still does, due to the increased duration of an item because of a migration. 

Our main result here is an algorithm for \dynd whose competitive ratio is the best-of-both worlds: it is as good as doing no migrations ($\mu$), but it can be much better if the delay cost is small ($\sqrt{C}$). 

\migrationDelay*

For the remainder of this section we present the algorithm that obtains the guarantees of \Cref{thm:alg-delay-cost}. The proof of the corresponding lower bound is provided in Appendix \ref{app:lower-bound-delays}.


\subsection{Algorithm}

Our algorithm is based on the following idea: For each item, we have a migration schedule with pre-determined migration times for this item. These migration times are chosen so we can charge the delay cost of an item to its duration. Our formal algorithm is given in \Cref{AlgDelayCost}. Concretely, the migration schedule we use performs each item's first migration $\sqrt{C}$ time after the item's arrival and then every $C + \sqrt{C}$ time after that. The algorithm maintains two instances: the small items instance $I_s$ for packing items until their first migration (if any), and the big items instance $I_b$ for packing items after their first migration has taken place. Our algorithm uses as a subroutine the \firstfit strategy that assigns each arriving item to the earliest-opened bin that can accommodate it; if no such bin exists, then it places this item in a new bin.

\begin{algorithm}
	\SetAlgoLined
	\DontPrintSemicolon
	Maintain two instances, one for \emph{small} items $I_s$ and one for \emph{big} items $I_b$. Initially, both instances are empty.\;
	Upon arrival of an item, assign it to a bin in $I_s$ using \firstfit.\;
	\If{an item has been in $I_s$ for exactly $\sqrt{C}$ time}{migrate it, placing it into a bin in the instance $I_b$ using \firstfit.}
	\If{an item has been in $I_b$ for exactly $C + \sqrt{C}$ time since its most recent migration}{migrate it, placing it into a bin in the instance $I_b$ using \firstfit.} 
	\caption{Algorithm for Delay Cost}\label{AlgDelayCost}
\end{algorithm}


\subsection{Analysis}

To prove that \Cref{AlgDelayCost} has the competitiveness stated in \Cref{thm:alg-delay-cost}, namely that it is $O(\min(\mu, \sqrt{C}))$-competitive, we show the following:

\begin{itemize}
    \item If the ratio $\mu$ is small, then we recover the known $O(\mu)$-guarantee of \firstfit~\cite{nonClairvoyantDynBP}.
    
     \item Otherwise, $\mu$ is large, so our goal is a $O(\sqrt{C})$-approximation by using migrations. We first show that the delay cost incurred by an item can be bounded with respect to its initial duration.

     \item Finally, we show \firstfit performs well on the sub-instances (subject to the additional costs caused by migration delays) generated by our algorithm by controlling the ``effective $\mu$'' actually experienced by \firstfit.
\end{itemize}

So we split our analysis in two cases based on which term achieves $\min(\mu, \sqrt{C})$.

\medskip
\noindent\textbf{Case 1: $\mu < \sqrt{C}$.}
In this case, our goal is an $O(\mu)$-approximation. Every item has duration $< \sqrt{C}$ (by our assumption that the min duration is $1$). Thus, every item is assigned to $I_s$ upon arrival and departs before its first migration checkpoint, $\sqrt{C}$. It follows our algorithm never migrates any items, and its total active time is exactly that of running just \firstfit on the full instance. We recall that the performance of \firstfit is dictated by the parameter $\mu$, as shown in~\cite{nonClairvoyantDynBP}.

\begin{thm}[\cite{nonClairvoyantDynBP}]\label{thm:first-fit}
	\firstfit is $O(\mu)$-competitive for dynamic bin packing to minimize the total active time, where $\mu$ is the ratio between the maximum item duration over the minimum item duration.
\end{thm}

We conclude, by the above theorem, that our algorithm is $O(\mu)$-competitive in this case.

\medskip
\noindent\textbf{Case 2: $\mu \geq \sqrt{C}$.}
In this case, our goal is an $O(\sqrt{C})$-approximation. We first bound the increase in duration for each item caused by its migrations. We say an item's \emph{initial duration} is the item's duration if it is never migrated (i.e. $d_i$). The item's \emph{delayed duration}, which we denote by $\tilde{d}_i$, is the duration taking into account any migration delays, i.e., $\tilde{d}_i = d_i + \#(\text{migrations of item $i$}) \cdot C$. We show that by the design of our migration schedule, we can charge the increase due to delay costs to some portion of the initial duration.

\begin{lemma}\label{lemma:duration-increase}
	An item with initial duration $d \in [M \cdot \sqrt{C}, (M+1) \cdot \sqrt{C})$ for integer $M = 0, 1, \dots$ has delayed duration $\tilde{d} \leq d + M \cdot C$. In particular, $\tilde{d} \le O(\sqrt{C}) \cdot d$.
\end{lemma}
\begin{proof}
         Let $m$ be the number of times this item is migrated. If $m=0$, i.e., the item is never migrated, its delayed duration will equal its initial duration, trivially giving the desired bound. So assume throughout that $m \ge 1$.
         

    

    Since the delayed duration of the item is $d + m \cdot C$, the desired result is equivalent to showing that the number of migrations $m$ is at most $M$. The main idea for showing this upper bound is to view the delay costs as follows: If an item is migrated -- incurring delay cost $C$ -- we view the next $C$ time units as paying for this delay. We view all other time spent as using up the initial duration $d$ of the item. Thus, we say that the \emph{remaining duration} of an item at a given instant is $d$ minus the time spent using up the initial duration until then.

    Now let $t_1,\ldots,t_m$ be the times the item is migrated,
and consider the $m$-many disjoint time intervals $[a, t_1), [t_1, t_2), \ldots, [t_{m-1}, t_m)$ induced by the migrations, where $a$ is the arrival time of the item. We claim that each of these intervals decreases the remaining duration of the item by $\sqrt{C}$. For the first interval $[a, t_1)$, note that it has length $\sqrt{C}$ by definition of \Cref{AlgDelayCost}, and the item is not migrated during this interval. Thus, we only use up the initial duration here, and in particular we decrease the remaining duration by $\sqrt{C}$ -- the length of this interval. Each subsequent interval begins with a migration and has length $C + \sqrt{C}$ by definition of \Cref{AlgDelayCost}. Thus, we spend the first $C$ units of time paying the delay cost, and the next $\sqrt{C}$ units decreasing the remaining duration, proving the claim. 
	
	Because $d < (M+1) \sqrt{C}$, there can be at most $m \leq M$ such intervals, since each one decreases the remaining duration by $\sqrt{C}$. This shows the desired upper bound on the number of migrations and concludes the proof of the lemma.  
\end{proof}

It remains to analyze the quality of the packing generated by \firstfit as we migrate. Let $I$ be the input instance. Note that \Cref{AlgDelayCost} runs two sub-instances of items: Upon arrival, we run \firstfit on the initial portion of the items prior to their first migration ($I_s$), then we  run \firstfit on every subsequent migration ($I_b$). Our strategy is to control the ``effective $\mu$'' for both sub-instances. To do so, we split each item of $I$ into a collection of sub-items using the times when \Cref{AlgDelayCost} migrates the item as the split points.

Concretely, given item $i \in I$, suppose \Cref{AlgDelayCost} migrates this item at times $t_1, \dots, t_m$. We define the sub-items of $i$ -- all with size $s_i$ -- as follows: The first sub-item is the \emph{small parts} of $i$ with arrival time $a_i$ and duration $t_1 - a_i$ if $i$ is ever migrated, and duration $d_i$ otherwise. All subsequent sub-items are the \emph{big parts} of $i$ with arrival time $t_j$ and duration $t_{j+1} - t_j$ for all $j = 1, \dots, m-1$. The final large part (corresponding to when the item finally departs) has arrival time $t_m$ and duration $\tilde{d}_i - t_m$, where $\tilde{d}_i$ is the delayed duration of $i$. Let $\tilde{I}_s$ and $\tilde{I}_b$ be the instances consisting of all small- and big parts, respectively. Note that the sub-items in $\tilde{I}_b$ accrue delay cost caused by previous migrations. 

Observe that we have defined $\tilde{I}_s$ and $\tilde{I}_b$ to be exactly the items (re-)packed by \firstfit into $I_s$ and $I_b$ respectively. Thus, the total active time of \Cref{AlgDelayCost} in $I_s$ is exactly $\firstfit(\tilde{I}_s)$, and analogously for $I_b$. Because \Cref{AlgDelayCost} uses disjoint bins for \firstfit on both sub-instances, the total active time of our algorithm is then:
\begin{align}
\ALG(I) = \firstfit(\tilde{I}_s) + \firstfit(\tilde{I}_b), \label{eq:instanceSplit}
\end{align}
where for any instance $I$, we define $\ALG(I)$ and $\firstfit(I)$ to be the total active time of running \Cref{AlgDelayCost} and \firstfit on that instance, respectively. We define $\OPT(I)$ and $\mu(I)$ analogously in the following.

It remains to relate \firstfit on each sub-instance with $\OPT(I)$. The easier of the two is the small instance, because the sub-items do not accrue any delay cost.

\begin{lemma}\label{lemma:small-instance}
	We have $\firstfit(\tilde{I}_s) \le O(\sqrt{C}) \cdot \OPT(I)$.
\end{lemma}
\begin{proof}
	Every sub-item in $\tilde{I}_s$ has duration in the range $[1, \sqrt{C}]$: The lower bound is by assumption, and the upper bound is because each item is either migrated away or departs within time $\sqrt{C}$. Thus, we have $\mu(\tilde{I}_s) \leq \sqrt{C}$.
 	Applying \Cref{thm:first-fit} gives 
	\[\firstfit(\tilde{I}_s) \le O(\mu(\tilde{I}_s)) \cdot \OPT(\tilde{I}_s) \le O(\sqrt{C}) \cdot \OPT(I),\] 
	where in the final we use the fact that $\OPT(\tilde{I}_s) \leq \OPT(I)$, because every sub-item in $\tilde{I}_s$ arrives at the same time with the same size and has only shorter duration than its corresponding item in~$I$.
\end{proof}

The big instance is more involved because we need to take the delay cost into account.

\begin{lemma}\label{lemma:big-instance}
	We have $\firstfit(\tilde{I}_b) = O(\sqrt{C}) \cdot \OPT(I)$.
\end{lemma}
\begin{proof}
	Note that every sub-item in $\tilde{I_b}$ has duration in the range $[C, C + \sqrt{C}]$: 
 The lower bound is because every sub-item in $\tilde{I_b}$ was created because of a migration, which guarantees that it will still be in the system for the next $C$ amounts of time (due to the delay cost), and also because we do not migrate the item (i.e., start the next sub-item) within the next $C$ times;
 the upper bound is because within the next $C + \sqrt{C}$ times after the creation of the sub-item, we will either migrate the item, starting a new sub-item, or the item departs from the system.
 It follows that $\mu(\tilde{I_b}) \leq \frac{C + \sqrt{C}}{C} \leq 2$.
	
	Thus, applying \Cref{thm:first-fit} gives
	$\firstfit(\tilde{I_b}) \le O(\OPT(\tilde{I_b})).$
	It remains to show that $\OPT(\tilde{I_b}) \le O(\sqrt{C}) \cdot \OPT(I)$. 
	
	We define the \emph{volume} and \emph{span} of an instance $I'$ by $\vol(I') = \sum_{i \in I'} s'_i \cdot d'_i$ and by $\spa(I') = | \bigcup_{i \in I'} [a'_i, a'_i + d'_i]\, |$, that is, the amount of times where there is at least one item in the system.
 Notice that these notions are based on the original instance, independent of the delays incurred by a given algorithm. These quantities are a good estimate of the optimal solution of \dyn~\cite{AzarVainstein}; since $\OPT$ for \dyn and \dynd are the same, because the latter accrues no delay cost, it also holds for our setting.
 
	\begin{claim}[\cite{AzarVainstein}, page 80]
		For any instance $I'$ of \dynd, we have $\OPT(I') = \Theta(\vol(I') + \spa(I'))$.
	\end{claim}
 
	Thus, our strategy is to relate the volume and span of $\tilde{I_b}$ with that of our original instance $I$. For the volume, we use \Cref{lemma:duration-increase}: all sub-items in $\tilde{I_b}$ corresponding to the same item in $I$ have disjoint lifetimes, and their durations add up to at most $O(\sqrt{C})$ times the initial duration of the item in $I$. Thus, we have $\vol(\tilde{I_b}) \le O(\sqrt{C}) \cdot \vol(I)$.

	It remains to upper bound $\spa(\tilde{I}_b)$ as a function of $\spa(I)$. To do so, we define the instance $\tilde{I}$, which consists of items $i \in I$ with the same arrival times and sizes as in $I$, but their durations are now their delayed duration, $\tilde{d}_i$. Notice that $\spa(\tilde{I}_b) \leq \spa(\tilde{I})$, and so we focus on upper bounding the right-hand side. 
	
	For that, consider a minimal sub-collection $\tilde{I}'$ of the items $\tilde{I}$ that achieves the same span. Because $\tilde{I}'$ is minimal, at any time there are at most two active items in $\tilde{I}'$ (if there were more, then we could keep the at most two items with the earliest arrival time and latest departure that intersect this time, contradicting the minimality). Thus, we can bound the span as
    \begin{align*}
        \spa(\tilde{I}) = \spa(\tilde{I}') = \bigg| \bigcup_{i \in \tilde{I}'} [a_i, a_i + \tilde{d}_i]\bigg| \le \sum_{i \in \tilde{I}'} \big|[a_i,a_i + \tilde{d}_i] \big| \le O\big(\sqrt{C}\big) \sum_{i \in \tilde{I}'} \big|[a_i,a_i + d_i] \big|,
    \end{align*}
    since again by \Cref{lemma:duration-increase} we have $\tilde{d}_i \le O(\sqrt{C}) \cdot d_i$. Since each time $t$ is counted at most twice in the intervals $[a_i, a_i + \tilde{d}_i]$ (and hence in the smaller intervals $[a_i, a_i + d_i]$), we have the ``reverse union bound''
    \begin{align*}
        \sum_{i \in \tilde{I}'} \big|[a_i,a_i + d_i] \big| \le 2 \bigg|\bigcup_{i \in \tilde{I}'} [a_i,a_i + d_i] \bigg| \le 2 \bigg|\bigcup_{i \in I} [a_i,a_i + d_i] \bigg| = 2\spa(I). 
    \end{align*}  
    Chaining the above inequalities gives:
	$\spa(\tilde{I}_b) \leq \spa(\tilde{I}) \leq O(\sqrt{C}) \cdot \spa(I),$
	as required.
	
	To conclude, we have $\OPT(\tilde{I}_b) \le O(\vol(\tilde{I}_b) + \spa(\tilde{I}_b)) = O(\sqrt{C}) \cdot (\vol(I) + \spa(I)) = O(\sqrt{C}) \cdot \OPT(I)$, and plugging this in to our bound on \firstfit gives $\firstfit(\tilde{I_b}) = O(\sqrt{C}) \cdot \OPT(I)$.
\end{proof}

Combining equation \eqref{eq:instanceSplit} with the upper bounds from \Cref{lemma:small-instance} and \Cref{lemma:big-instance} completes the proof of \Cref{thm:alg-delay-cost}.

\bibliography{ref}

\newpage
\appendix 
\noindent {\LARGE \bf Appendix}



\section{Lower Bound for \dyn with Linear Migration}\label{app:unit-migration-LB}

\migrationtradeoff*

    \begin{proof}
    We assume $\alpha \le \frac{1}{4}$, else the result is vacuously true. Consider the following random instance: Let $s$ be such that $\frac{1}{s} = \lfloor \frac{1}{4\alpha}\rfloor$, and $k \ge 1/s$ be a multiple of $s$. There are $n = k\cdot (1/s)$ items, each of size $s$. They are all released at time 0 and with durations $S_1,S_2,\ldots,S_n$ that are each independently set to $1$ (small) with probability $1-s$ and set to $\mu$ (big) with probability $s$, for a scalar $\mu > 1$ to be set later.

    First we compute $\OPT$. First, only $k$ bins are needed. Also, the optimal solution is to collocate the big items together in as few bins as possible, each of duration $\mu$, and then place the small jobs together on other bins. More precisely, since $1/s$ items fit in a bin, letting $X_j$ be the indicator that $S_j = \mu$ we see that $\OPT$ opens $\lceil \frac{\sum_j X_j}{1/s} \rceil \le s \cdot \sum_j X_j + 1$ bins to place the big items, and additional $\lceil \frac{n - \sum_j X_j}{1/s}\rceil \le s \, (n-\sum_j X_j) + 1$ bins to fit the remaining small items. The total cost of the solution is then at most $\mu \, (s \cdot \sum_j X_j + 1) + s \, (n-\sum_j X_j) + 1$. Since $\Pr(X_j = 1) = s$ and $n s = k$, the expected cost of this solution is at most 
    \begin{align}
        \mu \, (s k + 1) + s \, (n- k) + 1 \le \mu \, (s k + 1) + k + 1.  \label{eq:lbOPT} 
    \end{align}


    Now consider any randomized algorithm that makes at most a $\alpha n$ migrations in expectation; notice $\alpha n = \alpha \, \lfloor \frac{1}{4\alpha} \rfloor \, k \le \frac{k}{4}$.
    In fact, we consider more powerful algorithms that in expectation can \emph{delete} at most $\frac{k}{4}$ items at any point strictly after their arrival. We show that such algorithms have cost at least $\frac{k\mu}{4}$. 
    
    For that, let $J_i$ denote the set of items assigned by the algorithm to bin $i$ at the time of their arrival, and let $B_i$ be the event that bin $i$ receives a big item (i.e., there is an item $j \in J_i$ with $S_j = \mu$). Since the algorithm does not know the durations of the items when they are assigned, the assignments $J_i$'s are independent from the item durations $S_j$'s. Thus, we can condition on the assignment to bin $i$ and use the independence of the durations $S_j$'s to get that that conditional probability that bin $i$ does \emph{not} get a big item is
    \begin{align}
        \Pr(B^c_i \mid J_i) = \Pr\bigg(\bigwedge_{j \in J_i} (S_j = 1) \bigg) = \prod_{j \in J_i} \Pr(S_j = 1) = (1 - s)^{|J_i|} \le e^{- s |J_i|} \le 1 - \frac{s |J_i|}{2}, \label{eq:Bc}
    \end{align}
    where the second equation uses the independence of the $S_j$'s, the next to last inequality uses the estimate $(1-x) \le e^{-x}$, valid for all $x$; the last inequality uses the estimate $e^{-x} \le 1 - \frac{x}{2}$ valid for all $x \in [0,1]$ and the fact that $|J_i| \le 1/s$, since the size of the items ensures that at most $1/s$ of them can be assigned to a bin. Taking expectation over $J_i$, we obtain that $\Pr(B_i) \ge \frac{s \cdot \E |J_i|}{2}$. Since $\sum_i |J_i|$ equals the total number of items $n = k/s$, the expected number of bins that at time 0 (so no deletions have occurred) contain a big item is lower bounded as
    \begin{align}
        \E[\textrm{\#\,bins with a big item}] \ge \frac{s}{2} \cdot \E \sum_i |J_i| = \frac{k}{2}.  \label{eq:LBMigration}
    \end{align}
    Thus, even if the algorithm performs $\frac{k}{4}$ item deletions in expectation, in expectation at least $\frac{k}{4}$ bins remain with a big item until time $\mu$; this gives the algorithm expected cost at least $\frac{k \mu}{4}$, as claimed. 
    
    Thus, the competitive ratio of any such randomized algorithm is at least 
    $$\frac{k \mu / 4}{\mu \, (s k + 1) + k + 1. } = \frac{1}{4} \frac{1}{s + \frac{1}{k} + \frac{1}{\mu} + \frac{1}{k\mu}}.$$ Taking $k$ and $\mu$ to $\infty$ gives that the competitive ratio is at least $\frac{1}{4s} = \frac{1}{4} \lfloor \frac{1}{4\alpha} \rfloor$, concluding the proof (in fact, setting $\mu = k = \frac{1}{s}$ suffices to obtain this result with different constants).  
    \end{proof}

\section{Lower Bound for \dynd} \label{sec:lower-bound-delay} \label{app:lower-bound-delays}

We prove the following lower bound, which in particular implies a hardness of $\min\{\sqrt{C}, \mu\}$, where $\mu$ is the ratio of max over min duration of the items. 

\begin{lemma} \label{lemma:LBmig}
    Consider the \dynd problem. For any $C \ge 1$, there exists a family of instances with minimum item duration 1 and maximum duration $\mu = \sqrt{C}$ where every randomized algorithm is at least $\frac{\sqrt{C}}{30}$-competitive.
\end{lemma}

\begin{proof}
    Assume $\sqrt{C} \ge 30$, else the result trivially holds (i.e., the algorithm has to pay at least $\OPT$).

    We use the same instance as in \Cref{lem:migrationtradeoff} but with $k= \frac{1}{s} = 2\sqrt{C}$ and $\mu=\sqrt{C}$: there are $4C$ requests, each of size $1/2\sqrt{C}$. They are all released at time 0 and with durations $S_1,S_2,\ldots,S_{C}$ that are each independently set to $1$ (small) with probability $1-\frac{1}{2\sqrt{C}}$ and set to $\sqrt{C}$ (big) with probability $\frac{1}{2\sqrt{C}}$. Call this random instance $I$.

    The development in the proof of Theorem \ref{lem:migrationtradeoff} up to inequality \eqref{eq:lbOPT} gives that the optimal solution for $I$  has expected cost $$\E\, \OPT(I) \le 2 \mu + k + 1 \le 5 \sqrt{C}.$$
    
    We now analyze the cost of any online randomized algorithm.
    
    \begin{claim} \label{claim:alg}
        Any online randomized algorithm for \dynd has expected cost at least $\frac{C}{6}$ on the instance $I$. 
    \end{claim}

    \begin{proof}
    Notice that due to the delay cost, whenever the algorithm makes a migration its costs is at least $C$ (the migrated item will have duration at least $C$, so one bin needs to be kept open for it). Thus, if the algorithm makes some migration with probability at least $\frac{1}{3}$ when executing over $I$, the claim directly holds. So suppose the probability it makes a migration is less than $\frac{1}{3}$. 

    As in the proof of \Cref{lem:migrationtradeoff}, let $J_i$ be the set of items assigned by the algorithm to bin $i$ at the time of their arrival, and $B_i$ be the event that bin $i$ has a big item. 
    Inequality \eqref{eq:Bc} gives that $\Pr(B_i \mid J_i) \ge \frac{|J_i|}{4 \sqrt{C}}$ (recall that here we set $\frac{1}{s} = 2\sqrt{C}$); since the $|J_i|$'s always add up to the total number of items, $4C$, the expected number of bins with a big item conditioned on the assignment $(J_i)_i$ is $\E [\sum_i B_i \mid (J_i)_i] \ge \frac{1}{4 \sqrt{C}} \sum_i |J_i| = \sqrt{C}$. Moreover, since the $B_i$'s are independent conditioned on the assignment $(J_i)_i$ (they only depend on the duration of the items, which are independent of the assignment and independent of each other), employing the standard multiplicative Chernoff bound (for example, Theorem 1.1. of~\cite{concentration}) we have that $$\Pr\bigg(\sum_i B_i \,\le\, \frac{1}{2} \E\Big[\sum_i B_i \,\Big|\, (J_i)_i\Big] ~\bigg|~ (J_i)_i \bigg) \le e^{-\frac{\E \sum_i B_i}{8}} \le e^{-\frac{\sqrt{C}}{8}}.$$ Thus, for any assignment $(J_i)_i$, with probability at least $1 - e^{-\frac{\sqrt{C}}{8}}$, there are at least $\frac{1}{2} \E[\sum_i B_i\mid (J_i)_i] \ge \frac{\sqrt{C}}{2}$ bins with a big item at time 0 (i.e. before any migration is done).
    By taking a union bound, with probability at least $1 -  e^{-\frac{\sqrt{C}}{8}} - \frac{1}{3}$ this event holds and the algorithm does not make any migrations. In such scenarios, the cost of the algorithm is at least $\sqrt{C}$ times the number of bins with some big item. Thus, the cost of the algorithm is at least $\sqrt{C} \cdot \frac{\sqrt{C}}{2} = \frac{C}{2}$ with probability at least $1 -  e^{-\frac{\sqrt{C}}{8}} - \frac{1}{3} \ge \frac{1}{3}$, the last inequality using the assumption $\sqrt{C} \ge 30$. This concludes the proof.  
    \end{proof}
    
    Combing the upper bound $\E \OPT(I) \le 5 \sqrt{C}$ with the previous claim gives that every algorithm is at least $\frac{\sqrt{C}}{30}$-competitive. This concludes the proof of \Cref{lemma:LBmig}.
\end{proof}


\section{\dyn with Size-Cost Migrations} \label{app:size-migration}

In this section, we consider \dyn but where the migration cost is not the number of migrated items, but rather the total size of the migrated items (i.e. each time we migrate item $i$, we pay $s_i$ migration cost). We prove an analogous collection of results (\Cref{thm:sizeCost}) as for the unit cost case. Our first result is that sublinear migrations (in terms of total size of migrated items) gives no asymptotic advantage over zero migrations. To state our results, we say the \emph{total size} is the sum of all item sizes $\sum_i s_i$.

\begin{thm}\label{thm:sizesublinear}
	Every randomized algorithm for \dyn that migrates -- in terms of item sizes -- at most $o(\textrm{total size})$ in expectation has competitive ratio at least $\frac{\mu}{6}$.
\end{thm}

Second, for the regime where we migrate at most $\alpha \cdot (\textrm{total size})$ of item sizes for $\alpha \in (0,1)$, with the same idea of classifying bins as Bad or Good, we can obtain the same multiplicative trade-off between approximation and migrations as in \Cref{thm:unit-migration-costs}, but this time with only constant additive loss. Concretely, we show the following.

\begin{thm}\label{thm:sizecost}
	For every $\alpha \in (0,\frac{1}{2})$, there is an online algorithm for \dyn that migrates at most a $\frac{\alpha}{1-2\alpha}$-fraction of the total size of items and for every time $t$ has at most $\frac{1}{\alpha} \OPT_t + 1$ open bins. 
\end{thm}

Further, this trade-off is asymptotically optimal.

\begin{thm}\label{thm:sizetradeoff}
	For every $\alpha \in (0,1)$, any randomized algorithm for \dyn that migrates at most an $\alpha$-fraction of the total size of items in expectation has competitive ratio at least $\frac{1}{4} \lfloor \frac{1}{4\alpha} \rfloor$.
\end{thm}

As in the unit cost case, we can summarize the power of migrations for size costs in three regimes. Consider an algorithm that migrates at most $\alpha \cdot (\textrm{total size})$ of item sizes:\rnote{see new list here}

\begin{itemize}
	\item \textbf{(New)} If $\alpha \rightarrow 0$ as $n \rightarrow \infty$, then the algorithm has no advantage over zero migration algorithms.
	\item \textbf{(New)} If $\alpha \le 1$ is constant, then the correct trade-off between the number of migrations and competitive ratio is \emph{multiplicative} -- doing $\ell$-times as many migrations (in terms of item sizes) improves the competitive ratio from $\Theta(\frac{1}{\alpha})$ to $\Theta(\frac{1}{\ell \alpha})$.
	\item \textbf{(From \cite{feldkord2018fully})} If $\alpha > 1$, then the correct trade-off is \emph{additive} -- doing $\ell$-times as many migrations (in terms of item sizes) decreases the competitive ratio from $1 + O(\frac{1}{\alpha})$ to $1 + O(\frac{1}{\ell \cdot \alpha})$. Further, improving the competitive ratio requires increasing the number of migrations or additive term in the competitive ratio by a $poly(n)$-factor.
\end{itemize}

We now prove the three theorems.

\subsection{Sublinear Lower Bound}

We recall that \Cref{cor:sublinear} gives that any algorithm that migrates at most $o(n)$ items (i.e. sublinear unit cost) in expectation has competitive ratio at least $\frac{\mu}{6}$. The hard family of instances in this corollary is obtained by applying \Cref{thm:sublinear} to the finite family of instances constructed in \Cref{lem:basicLB}. One can check that all item sizes are identical over all instances of the finite family from \Cref{lem:basicLB}. Applying \Cref{thm:sublinear} concatenates a sequence of such instances, so we can conclude that the final hard family of instances has identical item sizes. Thus, any algorithm that migrates $o(n)$ items is exactly one that migrates $o(\textrm{total size})$ of item sizes, so in particular the same lower bound of $\frac{\mu}{6}$ holds against algorithms that migrate $o(\textrm{total size})$ of item sizes in expectation. This completes the proof of \Cref{thm:sizesublinear}.

\subsection{Upper Bound} \label{app:size-migration-UB}

\subsubsection{Algorithm}

Our algorithm, \Cref{AlgMigrationCostsSize}, is similar to \Cref{AlgMigrationCosts}, but with two main differences. First, it takes a single input parameter $\alpha$; a bin transitions to Good when its load reaches $1-\alpha$ (this corresponds to setting $f = 1- \alpha$ in  \Cref{AlgMigrationCosts}). Second, items with size larger than $\alpha$ get packed separately on dedicated bins. Finally, note than in this setting there is no need to split items into classes based on their sizes, so \Cref{AlgMigrationCostsSize} is our final algorithm. 

\begin{algorithm}
	\SetAlgoLined
	\DontPrintSemicolon
	Upon arrival of a new item $i$ with size $s_i$: \;
	\Begin{
		If $s_i \geq \alpha$, then assign $i$ to a dedicated bin by itself.\;
		Else if there is a Bad bin with load $\leq 1-s_i$, assign $i$ to this bin. If its load becomes $\geq 1 - \alpha$, label the bin as Good.\;
		Otherwise, if there are Good bins with load $\leq 1-s_i$, assign $i$ to any of these bins.\;
		Otherwise, open a new bin, assign $i$ to it, and label it as Bad if its load is $ \leq 1 - \alpha$ and Good otherwise.\;}
	
	Upon departure of an item $i$: \;
	\Begin{
		If $i$'s bin was Good and now is left with load $< \alpha$, migrate all of its remaining items using \firstfit with the bins ordered as follows: Bad bins, then Good bins, then new bins.
	}
    \caption{Algorithm for \dyn with Size-Cost Migrations (parameter $\alpha$)} \label{AlgMigrationCostsSize}
\end{algorithm}

\subsubsection{Analysis}

We show that \Cref{AlgMigrationCostsSize} satisfies the guarantees of \Cref{thm:sizecost}. The analysis is analogous to those of our previous algorithms \Cref{AlgMigrationCosts} and \Cref{AlgMigrationCostsClass} carried out in \Cref{sec:migCostAnalysis}.

\paragraph{Number of bins opened.} First observe that at each time there is at most one Bad bin. The proof is analogous to \Cref{claim:num-bad}: Suppose at some time there are two Bad bins. Consider the time that the second such Bad bin is opened. It must be the case that an item of size $< \alpha$ arrived at this time but did not fit in the currently-open Bad bin, so this Bad bin has load $> 1 - \alpha$. However, then this bin would have been labeled as Good, so this is a contradiction.

Thus, with the exception of at most one Bad bin, all other (dedicated and Good) bins maintain load at least $\alpha$ at all times. This means that 
	\[(\text{number of dedicated and Good bins at time $t$}) \leq \frac{1}{\alpha}(\text{total size of items at time $t$}) \leq \frac{1}{\alpha} \OPT_t.\] We then conclude that for any time $t$, our algorithm has at most $\frac{1}{\alpha} \OPT_t + 1$ open bins.

\paragraph{Size of Migrations.} Note that in order for the load of a bin to be migrated, we know that the bin has been filled up to at least $1-\alpha$ and then its load dropped to less than $\alpha$. As a result, the load that departed is at least $1-2\alpha$, while the load that needs to be migrated is at most $\alpha$. As a result, the fraction of migrated load is at most $\frac{\alpha}{1 - 2 \alpha}$ of the total size of the items, as desired. This concludes the proof of \Cref{thm:sizecost}.


\subsection{Lower Bound}

One can check that in the lower bound construction of \Cref{lem:migrationtradeoff} every item has the same size. Thus, migrating an $\alpha$-fraction of the items corresponds exactly to migrating an $\alpha$-fraction of the total size. We conclude, the same lower bound on the competitive ratio holds for randomized algorithms that migrate at most an $\alpha$-fraction of the total size. This gives \Cref{thm:sizetradeoff}.





\section{Randomized Lower Bound for \dyn without Migrations} \label{app:basicLB}

    \begin{lemma} \label{lem:basicLB}
    No randomized online algorithm for \dyn that does zero migrations is better than $\frac{\mu}{2}$-competitive.
    \end{lemma}

    \begin{proof}
    We consider the same instance as in Theorem \ref{lem:migrationtradeoff} with $k = \frac{1}{s}$: For some $k$ to be set later, there are $k^2$ items, each of size $1/k$. They are all released at time 0 and with durations $S_1,S_2,\ldots,S_{k^2}$ that are each independently set to $1$ (small) with probability $1-\frac{1}{k}$ and set to $\mu$ (big) with probability $\frac{1}{k}$. 

    The development in the proof of Theorem \ref{lem:migrationtradeoff} up to inequality \eqref{eq:lbOPT} gives that the optimal solution for this instance has expected cost at most $2 \mu + k + 1,$ and development up to inequality \eqref{eq:LBMigration} gives that the assignments of any randomized online algorithm makes, in expectation, at least $\frac{k}{2}$ bins have an item of duration $\mu$; thus the expected cost of such algorithm is at least $\frac{k \mu}{2}$.

    Therefore, the competitive ratio of such algorithm is at least $\frac{k \mu / 2}{2 \mu + k + 1}$; taking $k \rightarrow \infty$ gives that the algorithm is at least $\mu/2$ competitive, concluding the proof (using $k = \mu$ suffices to obtain the same result with a different constant). 
    \end{proof}

\end{document}